\documentclass[letterpaper]{JHEP3}

\usepackage[dvips]{graphicx}
\usepackage{epsfig,multicol,bbm}
\usepackage{amsmath}
\usepackage{appendix}

\def\lsim{\mathrel{\hbox{\rlap{\hbox{\lower4pt\hbox{$\sim$}}}\hbox{$<$}}}}

\makeatletter

\newcommand{\Rmnum}[1]{\expandafter\@slowromancap\romannumeral #1@}
\makeatother

\title{Gamma-Ray Constraints  on Maximum Cosmogenic Neutrino Fluxes and
UHECR Source Evolution Models}

\author{Graciela B. Gelmini\\
Department of Physics and Astronomy, UCLA, 475 Portola Plaza, Los
  Angeles, CA 90095, USA\\
Email: \email{gelmini@physics.ucla.edu}}

\author{Oleg Kalashev\\
INR RAS,\\
 60th October Anniversary pr. 7a,\\
 117312 Moscow,  Russia\\ 
E-mail: \email{kalashev@ms2.inr.ac.ru}}

\author{Dmitri  V. Semikoz\\
APC, College de France,\\
 11 pl. Marcelin Berthelot,\\
  Paris 75005, France\\
E-mail: \email{dmitri.semikoz@apc.univ-paris7.fr}}

\abstract{The dip model  assumes that the ultra-high energy cosmic rays (UHECRs) above 10$^{18}$ eV consist  exclusively of protons and  is consistent with the spectrum and composition measure by HiRes.  Here we present the range of cosmogenic neutrino fluxes  in the dip-model which are compatible with a recent determination of the extragalactic very high energy (VHE)  gamma-ray diffuse background derived from 2.5 years of Fermi/LAT data. We show that  the largest fluxes predicted in the dip model would be detectable by IceCube in about 10 years of observation and are  within the reach of a few years of observation with the ARA project. In the incomplete UHECR model in which protons are assumed to dominate only above 10$^{19}$ eV, the cosmogenic neutrino fluxes could be a factor of 2 or 3 larger. Any fraction of  heavier nuclei in the UHECR at these energies would reduce the maximum cosmogenic neutrino fluxes.  We also consider here special evolution models in which the UHECR sources are assumed to have the same evolution of either the star formation rate (SFR),  or the gamma-ray burst (GRB) rate, or the active galactic nuclei (AGN) rate in the Universe and found that the last two are disfavored (and in the dip model rejected) by the new VHE gamma-ray background. }

\begin{document}

\section{Introduction}

The origin and composition of the ultra-high energy cosmic rays (UHECRs) remain  open questions in spite of the experimental advances of recent years due to the Auger  and HiRes collaborations.

 The Auger collaboration~\cite{Auger} has found evidence in its measurements of air shower profiles, both in the depth of shower maximum  and its fluctuation, of an increasing dominance of heavy nuclei primaries at high energies~\cite{xmax},  starting with light nuclei at $3 \times 10^{18}$~eV and approaching iron at $4 \times 10^{19}$.   The interpretation of the air shower profiles depends, however, on the hadronic interaction  models used, which have uncertainties at large energies. Auger has also found  a correlation of events above $ 5.5 \times 10^{19}$ eV with the distribution of nearby extragalactic astrophysical sources, including an excess around the direction of Centauros A, the nearest AGN. The correlation of UHECR arrival directions with  the distribution of relatively close AGN within an angular distance of 3$^o$ has weakened with time~\cite{correlation}. This correlation is compatible with proton primaries, which would contradict the  heavy composition extracted from the Auger shower profiles. However, it has been argued that the  excess of events  in the direction of Centaurus A, the most significant contribution to the anisotropy in the Auger data,  could be explained by heavy nuclei primaries emitted from the Virgo cluster~\cite{Semikoz-Virgo}.   

No tendency towards a heavier composition at high energies is found in the HiRes~\cite{HiRes} data, which are compatible with the UHECR above 10$^{19}$~eV consisting primarily of   protons~\cite{Abbasi:2009nf}. HiRes data seem to indicate a change in composition from heavy to light at energies close to 5$\times 10^{17}$ eV~\cite{chem_HiRes}. 
However, HiRes does not find  a correlation of the  arrival directions  of UHECR above $4.0 \times 10^{19}$ eV or $ 5.7 \times 10^{19}$ eV with astrophysical sources at a distance smaller than 10$^o$~\cite{Abbasi:2010xt}.
  
   The spectrum and composition measured by HiRes are, thus, consistent with assuming that all UHECR above $10^{18}$ eV are due to  extragalactic protons. This is what is assumed in the ``dip model"~\cite{dip-model}. In the past we called this  the ``minimal UHECR model"~\cite{minimal-GKS}, because it allows to  fit the UHECR spectrum above 10$^{18}$ eV with   a minimum number of parameters. It is  usually called the ``dip-model", because  the ankle feature in the spectrum at $\sim 4 \times 10^{18}$ eV is interpreted as an absorption ``dip" in the spectrum due to energy losses by electron-pair production of the protons interacting with the cosmic microwave background (CMB).
 
 Both HiRes~\cite{Abbasi:2007sv} and  Auger~\cite{PAO-spectrum}   have  observed a suppression in the spectrum above $4 \times 10^{19}$~eV. For HiRes this is  consistent with being due to the photo-pion production of  protons
  interacting with the CMB, i.e. the GZK feature~\cite{gzk}.  The GZK process produces pions, the decay of which produces  both ``cosmogenic neutrinos"~\cite{bere} (from $\pi^{\pm}$)  and   photons (from  $\pi^0$), which we  call ``GZK photons". 
Although they share the same production mechanism, ultra-high energy (UHE) GZK photons 
and cosmogenic neutrinos are affected very differently by intervening backgrounds. 
The flux of UHE GZK photons is affected by the poorly known radio background and extragalactic magnetic fields which do not affect  neutrinos. These photons only reach us from less than 100 Mpc away, i.e. at
redshifts $z<0.02$ (see e.g.~\cite{GKS}). Auger has placed stringent limits on the fraction
of UHECRs that are photons~\cite{augersdphotonfractionlimit}.
Cosmogenic neutrinos do not interact during propagation and thus 
reach us from the whole production volume. Thus the flux of cosmogenic neutrinos arriving to Earth depends 
strongly on the evolution of the sources, which affect mostly the density of sources far 
away.

 The cosmogenic neutrinos have been extensively studied theoretically  since 1969~\cite{bere} onwards  (see e.g.~\cite{reviewGZKneutrinos,Semikoz:2003wv} and references therein), and constitute one of the main high energy signals expected in neutrino telescopes, such as IceCube~\cite{ICECUBE},  ANITA~\cite{ANITA},  the future KM3NeT~\cite{KM3NeT} and ARA~\cite{ARA} or space based observatories such as JEM-EUSO~\cite{JEM-EUSO}.
 They could also be observed by  the Pierre Auger  Observatory~\cite{Auger} and the Telescope Array~\cite{TA}.  We want here to see if  the highest cosmogenic neutrino fluxes compatible with all present data are within the reach of these experiments. 
The expected flux of cosmogenic neutrinos is higher if the UHECR consist of protons.  Heavy  nuclei  would  interact with radiation backgrounds and  break up into lighter nuclei and nucleons, only a fraction of which would be above the energy threshold for pion-production. Therefore  fewer UHE neutrinos (and  GZK photons)  would be produced. Thus we assume proton primaries and, to be consistent with this assumption, use the UHECR spectrum  measured by HiRes~\cite{HiRes-spectrum, Abbasi:2007sv}, which is also higher than the Auger spectrum (thus it also leads to  larger secondary neutrino fluxes because  of this reason). Thus we assume the  dip model~\cite{dip-model}, which is self consistent model of UHECR. Because we are interested in producing upper bounds to neutrinos fluxes, we also consider
the case in which protons only fit the HiRes spectrum above 10$^{19}$ eV, although this is  an incomplete model which does not explain the UHECR data below this energy. This model was also studied in recent evaluations of cosmogenic neutrino fluxes, Refs.~\cite{Berezinsky:2010xa}  and \cite{Ahlers:2010fw}, where it was found to produce the highest maximum fluxes.  Following  Ref.~\cite{Berezinsky:2010xa}  we call it the ``ankle model".

In this paper we  present  the  range of cosmogenic neutrinos in the dip and ankle models obtained by varying all relevant parameters  defining the primary astrophysical extragalactic nucleon fluxes  with which the HiRes data 
are well fitted above  either 10$^{18}$ eV or 10$^{19}$ eV, respectively.
Our code  uses the  kinetic equation approach, which  provides fast results for each of the set of parameters we use to predict each flux. This feature allows us to scan the  parameter space  to look for the range of predicted  fluxes. We already studied the expected range of GZK photons in the dip model~\cite{minimal-GKS}. Here we  study the expected range of cosmogenic neutrinos using a similar statistical method and taking into account new data, in particular the extragalactic  flux of photons below 1 TeV extracted from measurements of the Fermi Space Telescope ~\cite{Abdo:2010nz, Neronov:2011kg}.  The  secondary electrons, positrons and photons quickly cascade on the CMB, infrared and visible backgrounds  to lower  energies and generate a $\gamma$-ray background in the GeV-TeV energy range, at which  the Universe becomes transparent to
 photons~\cite{wdowczyk}. We explore in particular the constraint imposed on the neutrino production  models  by a recent evaluation~\cite{Neronov:2011kg}  of the diffuse extragalactic very high energy (VHE) gamma-ray background using 2.5 years of Fermi/LAT data.

\section{Calculation of  Neutrino and Photon  Fluxes}

We use a numerical code originally presented in Ref.~\cite{kks1999}  
to compute the  cascades produced in the intergalactic medium 
 by an homogeneous distribution of sources emitting originally
only protons. This is the numerical code of Ref.~\cite{GKS}, with a few
modifications.   It  uses the  kinetic equation approach and
calculates the propagation of  nucleons, stable leptons  and photons
using the standard dominant processes (see e.g. Ref.~\cite{reviews1}).
For nucleons, it takes into account single and multiple pion production,
$e^{\pm}$ pair  production by protons and neutron $\beta$-decays.
 For photons, it includes $e^{\pm}$ pair production,
inverse Compton scattering (ICS) and double $e^{\pm}$ pair production processes.
For electrons and positrons, it takes into account Compton scattering, triple
pair production and synchrotron energy loss on extra galactic magnetic fields
(EGMF).  The propagation of nucleons and electron-photon cascades  is calculated self-consistently. Namely, secondary (and higher generation) particles arising in all reactions are propagated alongside with the primaries.

At energies below the threshold for  pair production on the CMB, the electromagnetic cascade proceeds through interactions with  the Extragalactic Background Light (EBL) at infrared and optical frequencies.  Contrary to earlier models for the EBL until a few years ago which differ from each other, recent calculations starting with Ref.~\cite{Franceschini2008} and including Ref.~\cite{Kneiske2010} have similar results, as was shown in the recent detailed study of Ref.~\cite{Dominguez2010}.  We use the EBL spectrum of Ref.~\cite{Kneiske2010}. The extragalactic magnetic fields are taken here to be zero, which  for  protons above 10$^{18}$ eV and VHE $\gamma$-rays is equivalent to having any value of the field smaller than 10$^{-11}$ Gauss.  Extragalactic magnetic fields of this magnitude, 10$^{-11}$ Gauss or smaller, are required in simulations of large scale structure formation~\cite{Dolag:2004kp} and predicted in models of  generation of cosmological magnetic fields~\cite{Neronov:2009gh}. Notice that only magnetic fields much larger than these, of the order of nG, would be required to change considerably the electromagnetic cascade development, because of the synchrotron radiation loses of high energy electrons, into photons at low energies not detected by Fermi-LAT~\cite{Berezinsky:2010xa}.

 A precise simulation of the electron-photon cascade requires much more computational power than
the simulation of a hadronic cascade because the former is driven by much faster processes. 
This especially applies to ICS processes, whose interaction length  approaches 1 kpc for electron energies  below 10 TeV. At these energies,   the ICS energy loss  in each single interaction  is at least two orders of magnitude smaller than the energy of the particle. Thus, we use a continuous energy loss approximation for ICS. This  approximation allows us to decrease the calculation time while controlling the accuracy of the simulation, which we also do by checking the  conservation  of energy in the cascade. The latest version of the numerical code used in this paper shows $1.5\%$ accuracy in terms of energy conservation.

We parametrize the initial proton flux emitted at all sources with the  power law,
\begin{equation}
F(E) = f~ \frac{1}{E^\alpha}~  \exp(-E/E_{\rm max} - E_{\rm min}/ E)~.
\label{proton_flux}
\end{equation}
The index $\alpha$ and  the minimum and maximum energy, $E_{\rm min}$ and  $E_{\rm max}$, are free parameters. We take $\alpha$ in the range 2 to 2.7 and  $E_{\rm max}$ between  $10^{20}$ eV  and $10^{21}$ eV. Values of $E_{\rm max} < 10^{20}$ eV do not fit well the HiRes spectrum.  For the dip model we take $E_{\rm min} = 10^{17}$ eV, safely below  the lowest energy $E_{\rm fit}=10^{18}$ eV  at which we fit the UHECR spectrum. For the ankle model, we take  $E_{\rm min} = 10^{18}$ eV  and fit the  UHECR spectrum above $E_{\rm fit}=10^{19}$ eV. The amplitude $f$ is fixed by fitting the final proton flux from all sources to the observed flux of UHECR~\cite{HiRes-spectrum, Abbasi:2007sv} above $E_{\rm fit}$,  and taking the UHECR flux below $E_{\rm fit}$ and the measured VHE $\gamma$ ray flux as upper bounds (to the proton and photon fluxes respectively) with the procedure explained below.

We assume several forms for the comoving source 
density or luminosity evolution with the redshift $z$.  Unless otherwise specified
 we assume the form
\begin{equation}
n(z)=n_0 (1+z)^m \theta(z_{\rm max}-z) \theta(z-z_{\rm min}) 
\label{evolution-function}
\end{equation}
for the comoving number density $n(z)$  of sources at redshift $z$ as function of  the present number density $n_0$. We take the parameter $m$  to be between 0 and 5 ($m=0$ corresponds to no evolution),  the  maximum distance to the sources to be $z_{max}=2$,  and  the minimum distance to the sources to be $z_{\rm min}=0$ (i.e. comparable with the interaction length). Notice that taking a relatively low  value  of $z_{max}$ is compensated by allowing for large values of the evolution parameter  $m$ (even 5) which considerably increase the source luminosity at large $z$.

\begin{figure}[h]
\begin{center}
\includegraphics[height=0.50\textwidth,clip=true,angle=270]{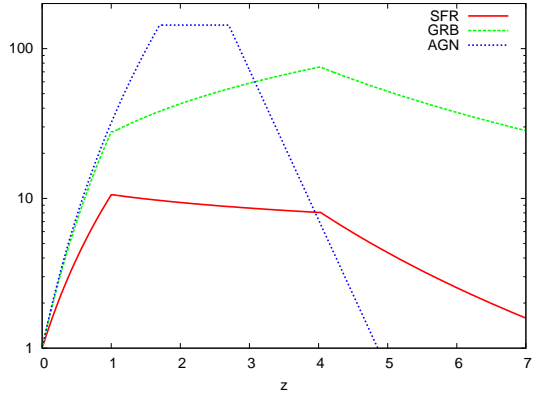}
\end{center}
\caption[...]{Particular evolution functions $S(z)$ in Eq.~\ref{S(z)} assuming  that UHECR sources have the evolution of either the star formation history (SFR), or the gamma-ray burst rate (GRB) or the active galactic nuclei rate (AGN).}
\label{evolution}
\end{figure}
  We also consider separately three particular source evolution models in which astrophysical sources of UHECR are assumed to have the  evolution functions $S(z)$ of either the star formation rate (SFR),  or the gamma-ray burst (GRB) rate, or the active galactic nuclei (AGN) rate in
the Universe. In this case, 
\begin{equation}
n(z)=n_0 S(z).
\label{S(z)}
\end{equation}
As in Ref.~\cite{Aharonian} for SFR we take~\cite{Yuskel-2008}
\begin{equation}
S_{\rm SFR}(z)= \left \{
\begin{array}{lll}
(1+z)^{3.4}, \,\,\,\, z<1 \\
(1+z)^{-0.3}, \,\,\,1<z<4\\
(1+z)^{-3.5}. \,\,\,\,z>4
\end{array} \right.
\label{SFR}
\end{equation}
for GRB we take $S_{\rm
GRB}(z)\propto (1+z)^{1.4}S_{\rm SFR}$~\cite{Yuskel-2007}, thus
\begin{equation}
S_{\rm GRB}(z)\propto \left \{
\begin{array}{lll}
(1+z)^{4.8}, \,\,\, z<1 \\
(1+z)^{1.1}, \,\,\,1<z<4\\
(1+z)^{-2.1}. \,\,\,z>4
\end{array} \right.
\label{GRB}
\end{equation}
and for AGN we take the function~\cite{Hasinger-2005,  Ahlers:2009rf}
\begin{equation}
S_{\rm AGN}(z)\propto \left \{
\begin{array}{lll}
(1+z)^{5.0},  \,\,\,z<1.7 \\
{\rm constant}, \,\,\,1.7<z<2.7\\
10^{(2.7-z)}. \,\,\,\,z>2.7
\end{array} \right .
\label{AGN}
\end{equation}
These evolution functions $S(z)$ are shown in Fig.~\ref{evolution}.
The redshift-distance relation is computed with the usual values for the cosmological parameters: 
$H=70$~km~s$^{-1}$~Mpc$^{-1}$  for the  Hubble constant,  $\Omega_{\Lambda}= 0.7$ for the dark energy density (in units of the critical density)  and $\Omega_{\rm m}=0.3$ for the dark matter density.  

As upper limit to the predicted VHE gamma-ray fluxes we use both the First-Year Fermi extragalactic  diffuse background~\cite{Abdo:2010nz} and a calculation of the background in Ref.~\cite{Neronov:2011kg}  which is lower than the first one in the 10-400 GeV energy range. Both fluxes are shown respectively in black and red in Figs.~\ref{BerezCompare},  \ref{SarkarCompare} and \ref{F4}.
 The analysis Ref.~\cite{Neronov:2011kg} using 2.5 years of the  Fermi/LAT  telescope, was done at high galactic latitudes $|b|>60^\circ$ using the same galactic model of the Fermi analysis. The estimation of the remaining diffuse cosmic ray flux was done with the prescription of the Fermi analysis, with the difference that the estimation of the suppression of the gamma-ray flux of the events classified as class 4  (those which are most confidently identified with $\gamma$-rays) was done independently of the  analysis of the signal from known blazars. With this analysis the diffuse background at energies   $E>10$ GeV  has a different spectrum than that of  the First-Year Fermi analysis~\cite{Abdo:2010nz}. The new spectral shape as function of energy closely follows the stacked spectrum  of BL Lac objects, which dominate the point source contribution at largest energies.
As we show below, the use of the VHE $\gamma$-ray flux in Ref.~\cite{Neronov:2011kg}  instead of the original First-Year Fermi background makes a  significant difference in terms of the largest cosmogenic neutrino fluxes allowed.

\section{Comparison with Other Calculations}

\begin{figure}[h]
\begin{center}
\includegraphics[height=0.60\textwidth,clip=true,angle=270]{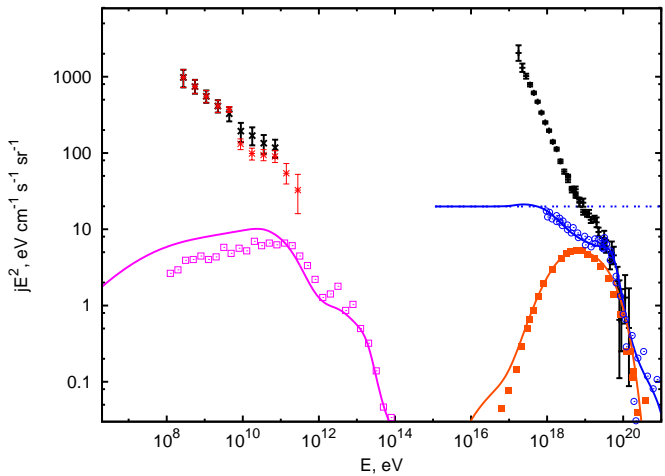}
\end{center}
\caption[...]{Comparison of proton, neutrino   (summed over flavors) and photon fluxes obtained with our code (shown respectively as blue, red and magenta continuous lines) with those presented  in the  Fig.~2 of Ref.~\cite{Berezinsky:2010xa}, shown  respectively as blue empty circles, red filled squares and magenta empty squares (see text for details). The HiRes UHECR spectrum (in black) and also the First-Year Fermi~\cite{Abdo:2010nz} (in black) and new lower~\cite{Neronov:2011kg} (in red) VHE $\gamma$-ray spectra are also  shown. The double-dotted blue line shows the original proton spectrum.}
\label{BerezCompare}
\end{figure}

\begin{figure}[h]
\begin{center}
\includegraphics[height=0.60\textwidth,clip=true,angle=270]{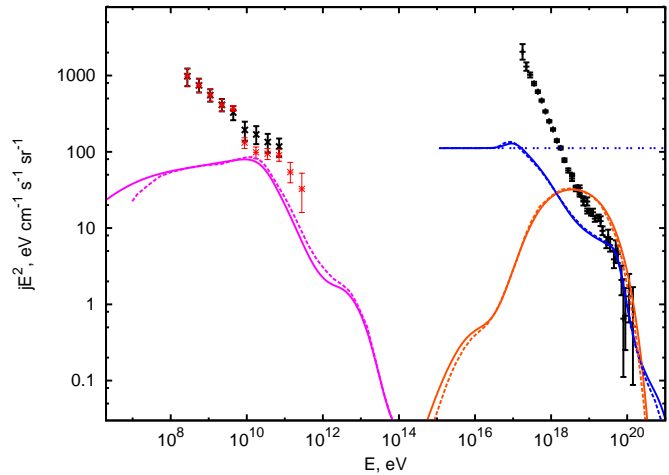}
\end{center}
\caption[...]{Comparison of proton (in blue), neutrino summed over all  flavors (in red) and photon (in magenta) fluxes obtained with our code, shown with continuous lines, with those presented  in the  Fig. B.7 of Ref.~\cite{Ahlers:2010fw}, shown with dashed lines  (see text for details). The HiRes UHECR spectrum (in black) and also the First-Year Fermi~\cite{Abdo:2010nz} (in black) and new lower~\cite{Neronov:2011kg} (in red) VHE $\gamma$-ray spectra are also shown.  The double-dotted blue line shows the original proton spectrum.}
\label{SarkarCompare}
\end{figure}

In Figs.~\ref{BerezCompare} and ~\ref{SarkarCompare} we compare the results of our program with those of other recent  calculations of cosmogenic neutrino and VHE $\gamma$-ray fluxes.

 In Ref.~\cite{Berezinsky:2010xa} the cascade evolution was computed both analytically  and with a Monte Carlo simulation program~\cite{Kachelriess:2004pc}, assuming a pure-proton composition model described by the parameters $\alpha$, $E_{\rm max}$, $m$ and $z_{\rm max}$ defined  as in Eqs.~\ref{proton_flux} and \ref{evolution-function}  above. In  Fig.~\ref{BerezCompare} we compare the photon and neutrino fluxes summed over flavors obtained with our code with those presented  in the  Figure 2 of Ref.~\cite{Berezinsky:2010xa}.  We use the same parameters defining the emission model, namely $\alpha=2$, $E_{\rm max}=10^{21}$ eV,  $z_{\rm max} = 2$ and  $m = 0$ in Eq.~\ref{evolution-function} and the same EBL model~\cite{Kneiske:2003tx} as in Ref.~\cite{Berezinsky:2010xa}.  The discrete  points  taken from Ref.~\cite{Berezinsky:2010xa} and reproduced in Fig.~\ref{BerezCompare} were obtained with the Monte Carlo method. 
   With the chosen values of the parameters the resulting proton spectrum does not fit well the HiRes data, thus, we  just normalized the  proton flux to the HiRes spectrum above $2\times10^{19}$ eV. Fig.~\ref{BerezCompare} shows that the proton and neutrino fluxes are in moderate agreement but the VHE photon flux predictions in the energy range measured by Fermi/LAT differ roughly by factor of $2$. 

Ref.~\cite{Ahlers:2010fw} also studied all-proton models of UHECR using an analytic model for the cascade evolution and fitted the resulting proton spectrum to the HiRes data. The injection spectrum Ref.~\cite{Ahlers:2010fw}  is again characterized by  the parameters $\alpha$, $m$, $E_{\rm max}$ and $E_{\rm min}$, as  in Eqs.~\ref{proton_flux} and \ref{evolution-function} above.
In  Fig.~\ref{SarkarCompare} we compare the results of our photon flux and neutrino flux (summed over flavors) calculations with those presented in Figure B.7 of Ref.~\cite{Ahlers:2010fw} with the same parameters, i.e. $\alpha=2$,  $E_{\rm max}=10^{21}$ eV, $m = 3$ and $z_{\rm max}=3$. These are parameters that do not fit well the HiRes spectrum and are use just for the purpose of comparing results.  We also use  the same EBL model\cite{Franceschini2008}  as in~\cite{Ahlers:2010fw}. Fig.~\ref{SarkarCompare} shows that our predictions for photon and neutrino fluxes are in very good agreement with those  of Ref.~\cite{Ahlers:2010fw}.

 \section{Range of Cosmogenic Neutrino Fluxes in the Dip Model of UHECR}  
 
 In this section we present the range of cosmogenic neutrino fluxes predicted by the dip model computed with  a method similar to that described  in Ref.~\cite{minimal-GKS}, but with some differences. In particular we use two goodness of fit tests, one using 
Pearson's chi-square and the other using the Poisson likelihood function.
\begin{figure}[ht]
\begin{center}
\includegraphics[height=0.32\textwidth,clip=true,angle=0]{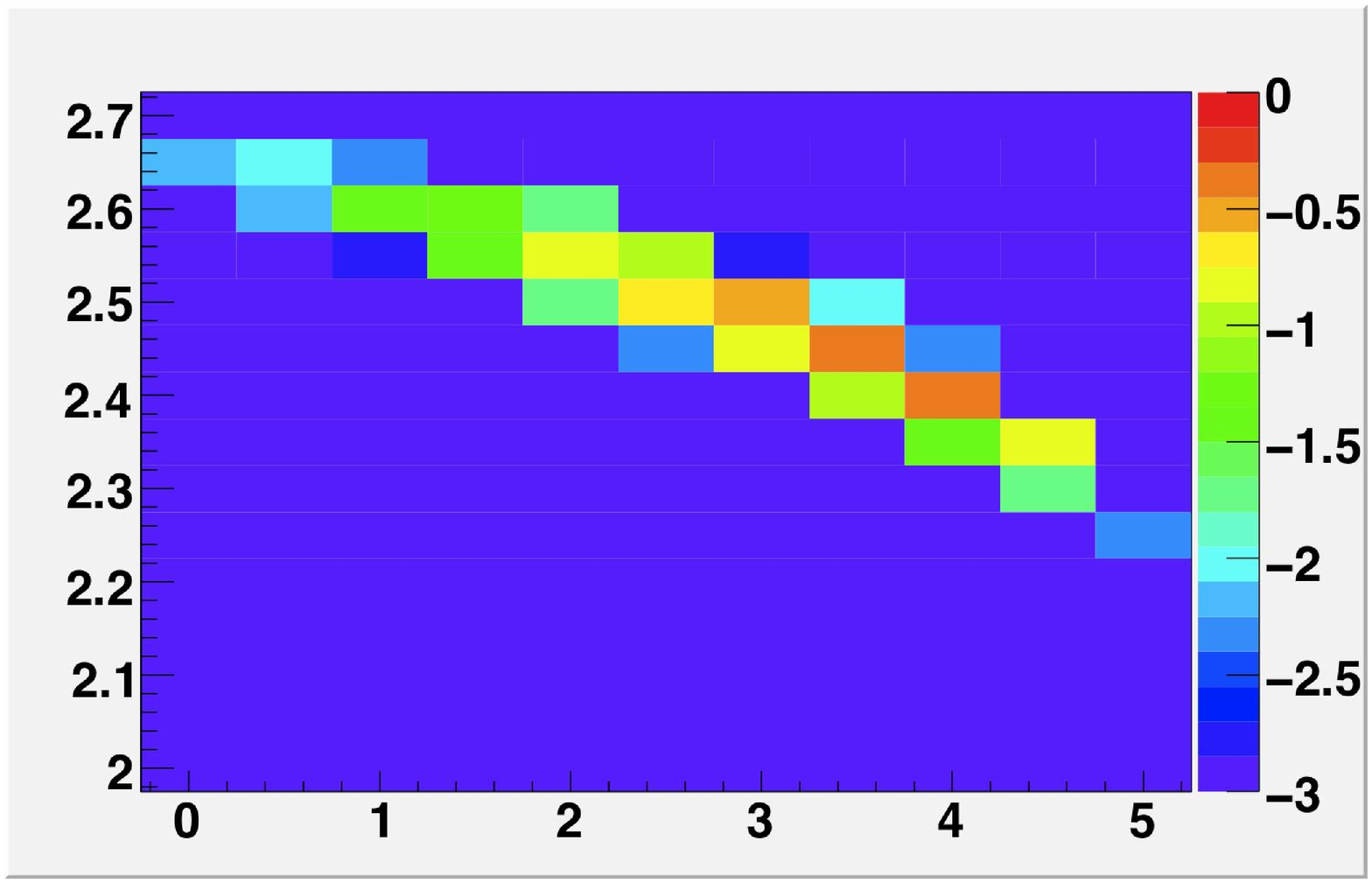}
\includegraphics[height=0.32\textwidth,clip=true,angle=0]{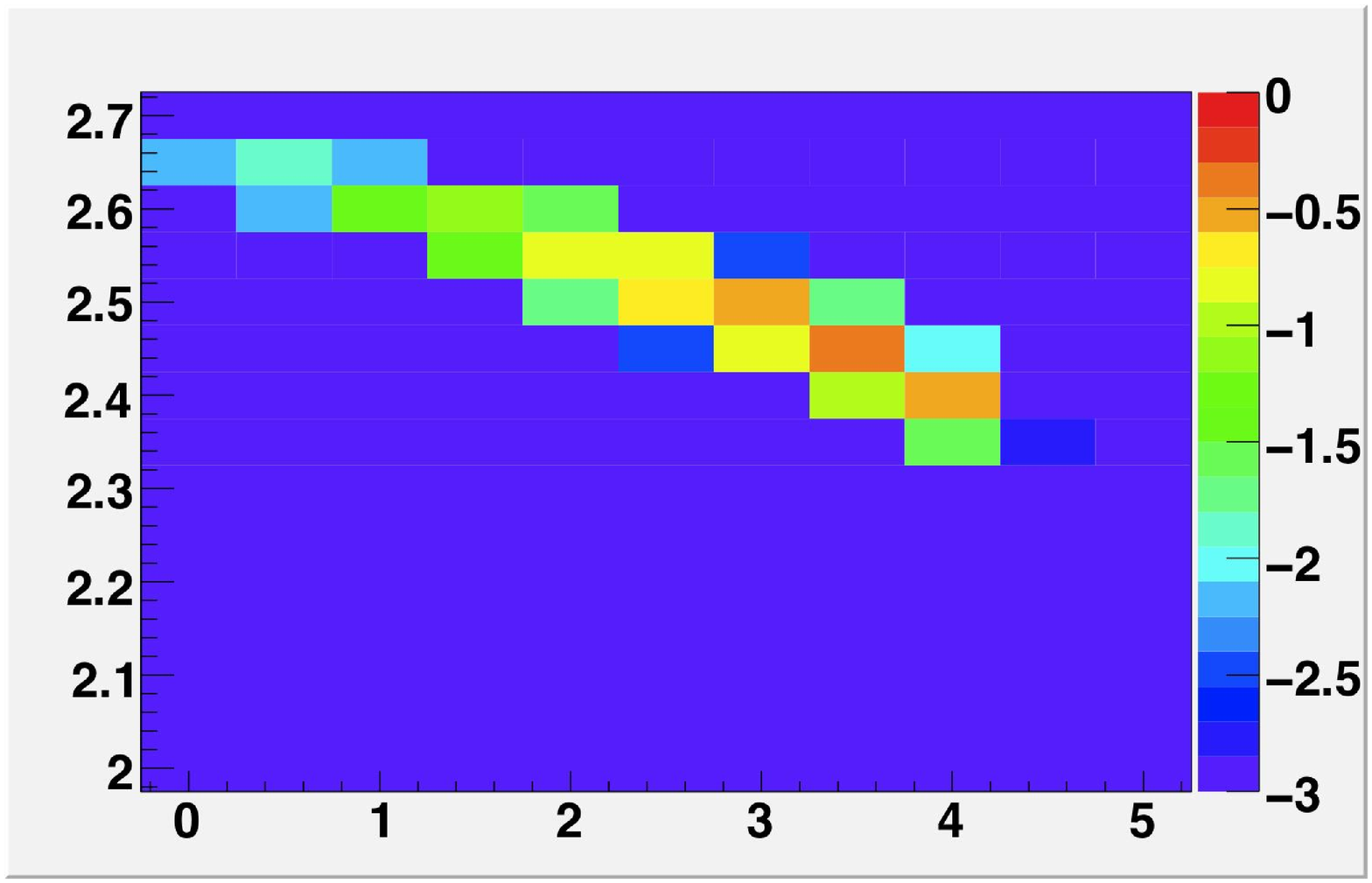}
\includegraphics[height=0.32\textwidth,clip=true,angle=0]{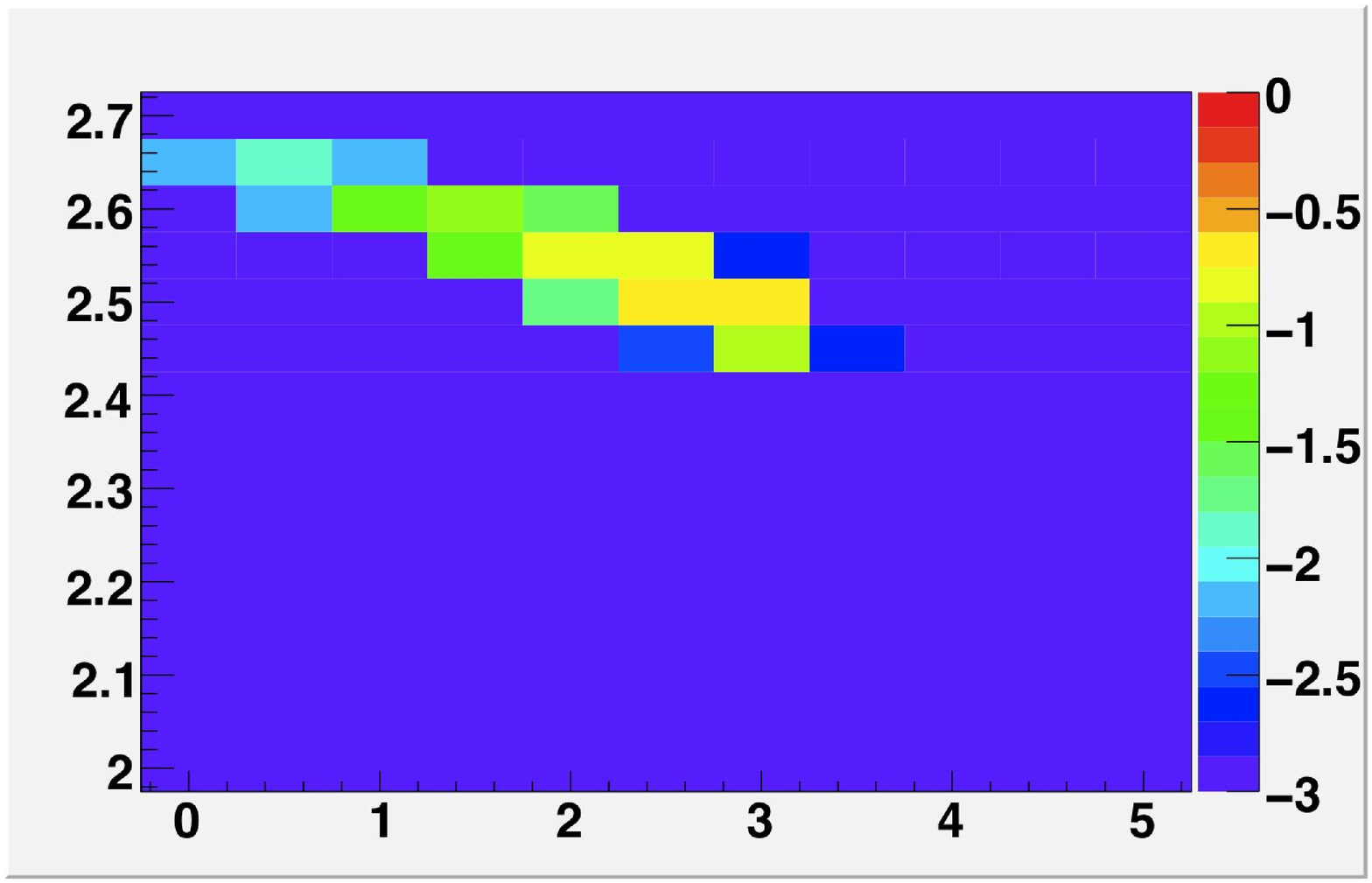}
\end{center}
\caption[...]{Consistency level of the predicted UHECR proton flux
with the HiRes spectrum  in the dip model and  \ref{F1}.a- (upper panel) no  upper bounds on the VHE $\gamma$-ray  fluxes imposed,   \ref{F1}.b -(middle panel)  the First-Year Fermi spectrum~\cite{Abdo:2010nz}  used as upper limit on the predicted VHE $\gamma$-ray  fluxes and  \ref{F1}.c -(lower panel) the spectrum of Ref.~\cite{Neronov:2011kg} used as upper limit  on the predicted VHE $\gamma$-ray  fluxes. The models in this figure have $E_{\rm min}=10^{17}$~eV, $E_{\rm max}=10^{21}$~eV,  $m$ from 0 to 5 in 0.5 steps and $\alpha$ from 2.00 to 2.70 in 0.05 steps. Color coded logarithmic chi-square $p$-value scale, from best ($p=1$) to worse ($p$ close to zero). Only models with log($p$)$> -1.30$ are accepted.}
\label{F1}
\end{figure}
 \begin{figure}[h]
 \begin{center}
  \includegraphics[height=0.60\textwidth,clip=true,angle=270]{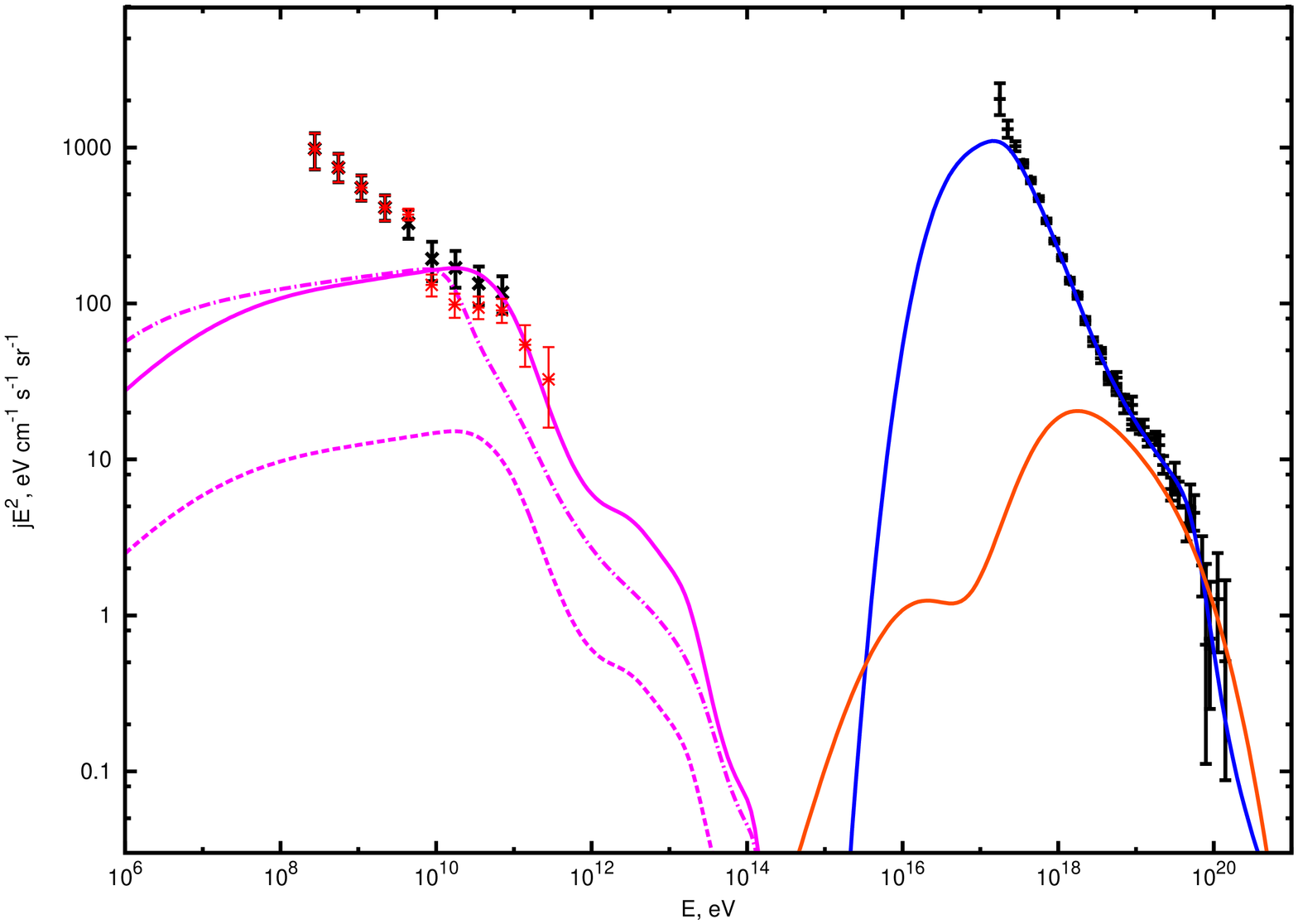}
\end{center}
\caption[...]{Fluxes of primary protons (blue), secondary photons (magenta) and neutrinos summed over flavors (red)
 for the dip model with $E_{\rm min}=10^{17}$~eV,  $E_{\rm max}=10^{21}$~eV, $\alpha=2.45$ and $m=3.5$, which is the best (largest $p$-value)  model of Fig.~\ref{F1}.b (with the First-Year Fermi VHE $\gamma$-ray spectrum, shown in black in the figure) and is forbidden in  Fig.~\ref{F1}.c (with the spectrum of Ref.~\cite{Neronov:2011kg}, shown in red in the figure). The dot-dashed magenta line shows the lower VHE photon flux predicted using the older evaluation of the EBL of Ref.~\cite{Stecker:2005qs}. The HiRes UHECR spectrum is also shown (in black). The flux of photons produced  only in the GZK process is  also shown (dotted line), which shows that the majority of neutrinos come from pair production.}
\label{F4}
\end{figure}   

Our procedure is the following. Among  all  models  characterized by the emission spectrum parameters and source evolution models defined  above we choose those which: 

1)- produce proton fluxes which fit the  HiRes spectrum~\cite{HiRes-spectrum, Abbasi:2007sv}  at energies $E > E_{\rm fit}$ (for the dip model $E_{\rm fit}=10^{18}$ eV),  taking into account the empty bins in the HiRes data above the highest energy UHECR events observed,  

2)-  produce proton fluxes not exceeding the HiRes spectrum at $E < E_{\rm fit}$ and 

3)- produce VHE $\gamma$-ray fluxes not exceeding the measured background assumed, 

\noindent at the 95\% confidence level.  We choose the value of the parameter $f$ in Eq.~\ref{proton_flux}, i.e. the
 amplitude of the injected spectrum, by minimizing  the combined  $\chi^2$. As just mentioned, the HiRes spectrum below   $E_{\rm fit}$ and the measured extragalactic VHE gamma-ray fluxes are taken as upper limits to the predicted proton and gamma-ray fluxes respectively.  Namely, if the predicted  proton fluxes below $E_{\rm fit}$ and VHE gamma-ray fluxes exceed their respective measured values, they are included in the calculation of the  $\chi^2$. If not, they are not included (so the number of data points in the calculation changes in principle with each model).  In this step we combine the high energy cosmic ray bins with number of event smaller than 5 into bins containing more than 5 events to ensure that  Pearson's  $\chi^2$ statistics is valid. We keep only models with a $p$-value $p>0.05= 10^{-1.30}$.  For these models, using the same fixed value of $f$ we calculate separately the goodness of fit for the bins with small number of events. Namely, we compute the Poisson likelihood function for the given value of $f$.  We then compute using a Monte Carlo technique the goodness of the fit or $p$-value defined as the fraction of  generated hypothetical experiments (observed spectra) with the same average number of evens (i.e. the predicted number) in each bin which results in a worse, namely a lower Poisson likelihood than the original one.  This procedure  for large number of events in each bin is equivalent to taking the $\chi^2$ distribution  without free parameters.  Only if the second $p$ value obtained in this way is also larger than 0.05 the model is accepted (notice that a higher $p$ value corresponds to a better fit, since more  hypothetical experimental results would yield a worse fit than the  one we obtained).  In this way we eliminate those models which are inconsistent  with the HiRes observed spectrum above $E_{\rm fit}$  and upper limits on the UHECR  flux below this energy and the VHE $\gamma$-ray flux  at the 95 \% C.L.
\begin{figure}[ht]
\includegraphics[height=0.50\textwidth,clip=true,angle=270]{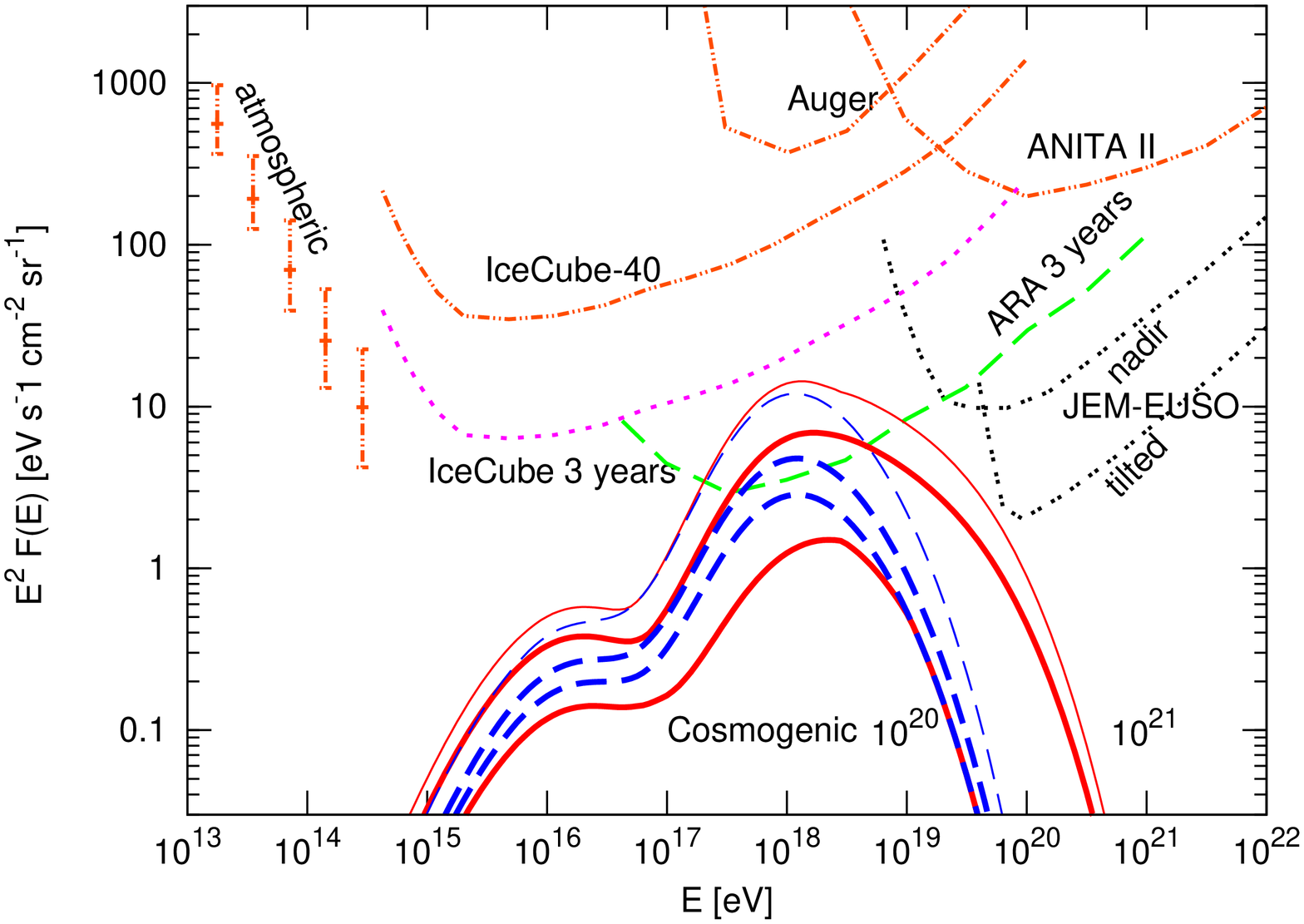}
\includegraphics[height=0.50\textwidth,clip=true,angle=270]{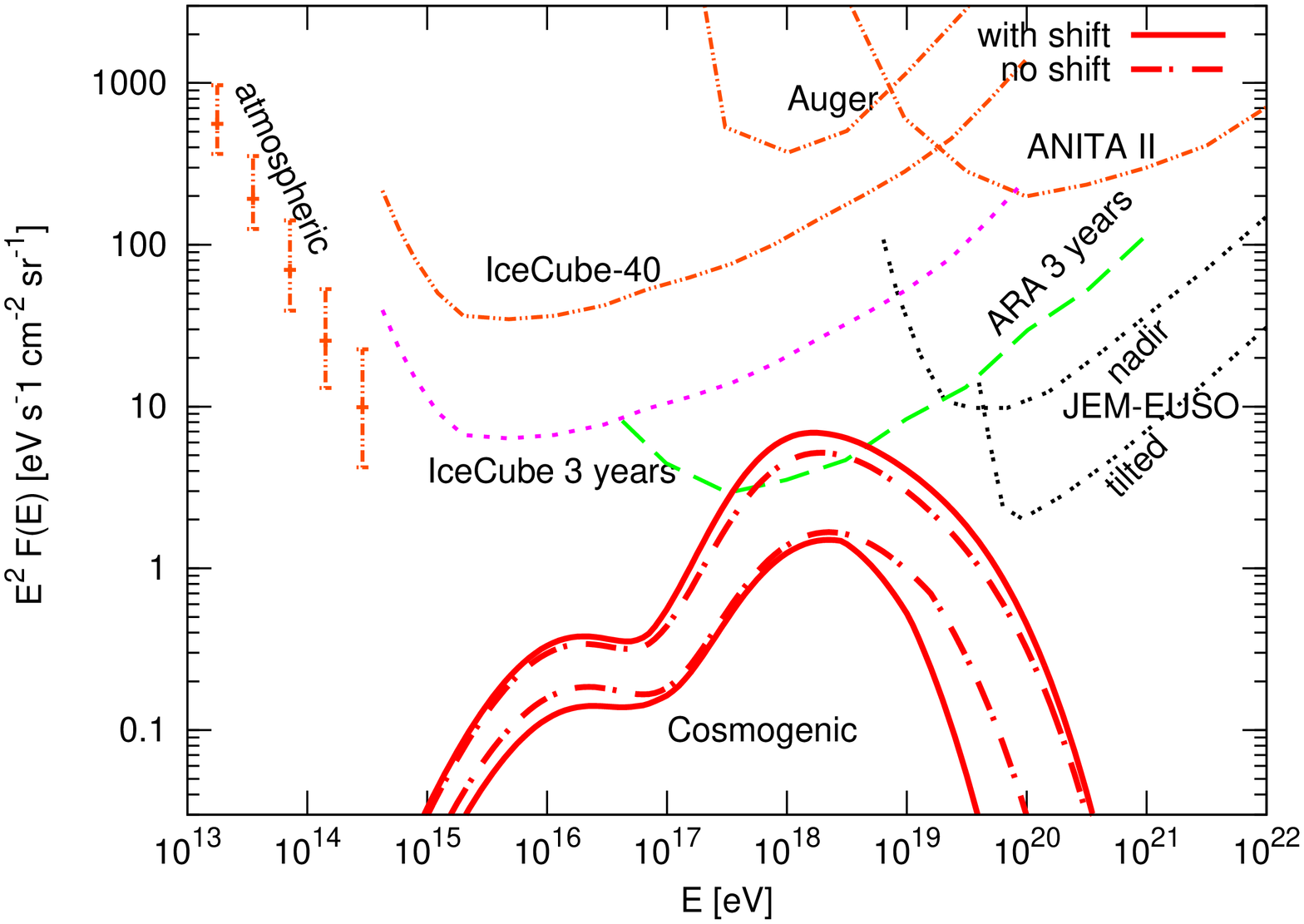}
\caption[...]{Maximum and minimum cosmogenic neutrino fluxes  averaged over flavors expected in the dip-model  (i.e. $E_{\rm min} = 10^{17}$ eV and $E_{\rm fit} = 10^{18}$ eV) using the evolution model of Eq.~\ref{evolution-function}. Thick and thin lines correspond to the VHE $\gamma$-ray spectrum of Ref.~\cite{Neronov:2011kg} and First-Year Fermi, respectively. \ref{F6}.a (left panel) for $E_{\rm max}=10^{20}$ eV (blue lines)  or  all $E_{\rm max}$ values, i.e. $10^{20}$ eV$\leq E_{\rm max} \leq 10^{21}$ eV (red lines),  including models with the energy scale of HiRes shifted by a factor in the range (1.30)$^{-1}$ to 1.30. \ref{F6}.b (right panel). For all $E_{\rm max}$ and the VHE $\gamma$-ray spectrum of Ref.~\cite{Neronov:2011kg} only, we compare the neutrino flux range obtained by shifting the HiRes energy scale (same as in the left panel) with the  range obtained without any  energy shift.
}
\label{F6}
\end{figure}      

\begin{figure}[ht]
\includegraphics[height=0.50\textwidth,clip=true,angle=270]{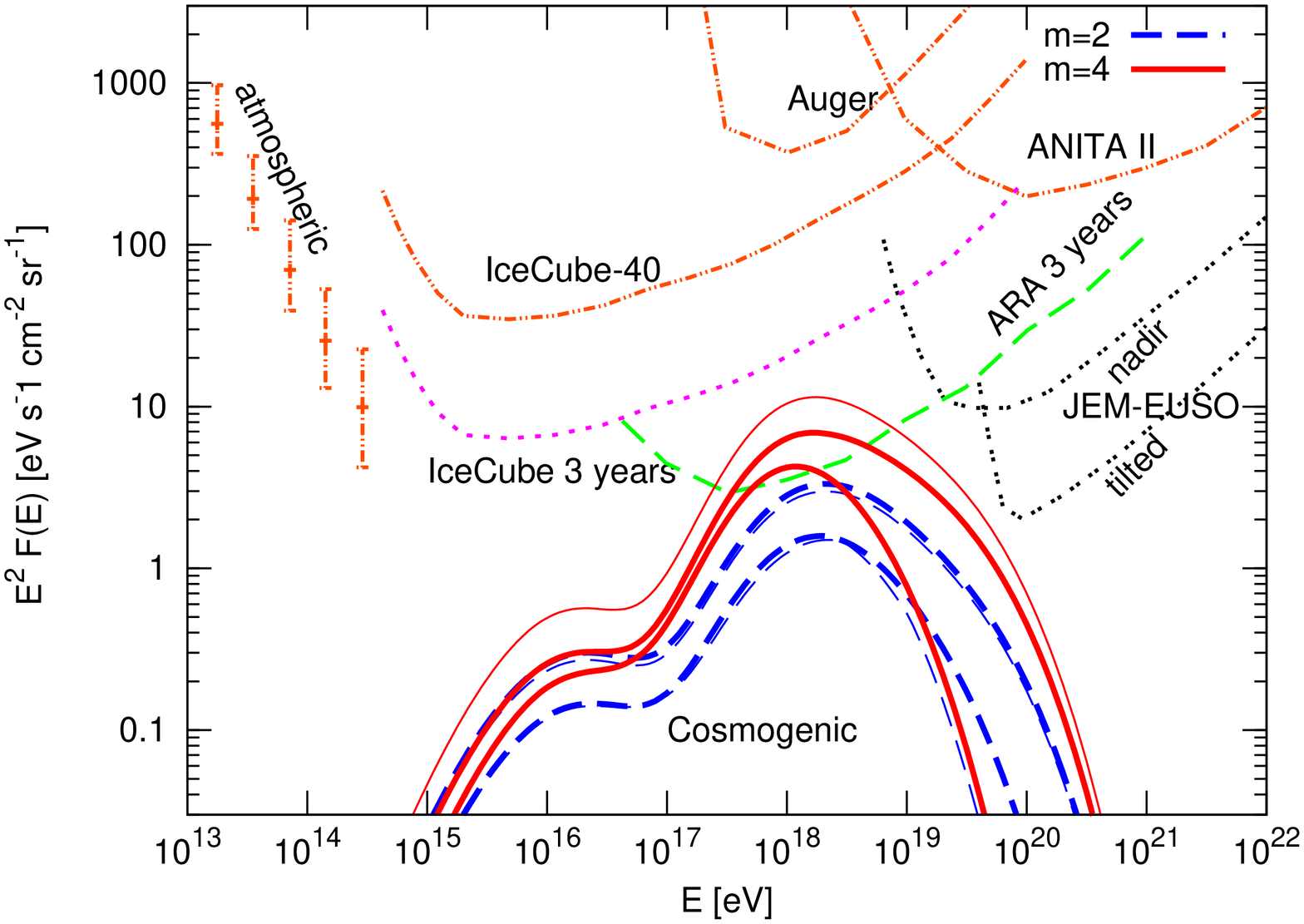}
\includegraphics[height=0.50\textwidth,clip=true,angle=270]{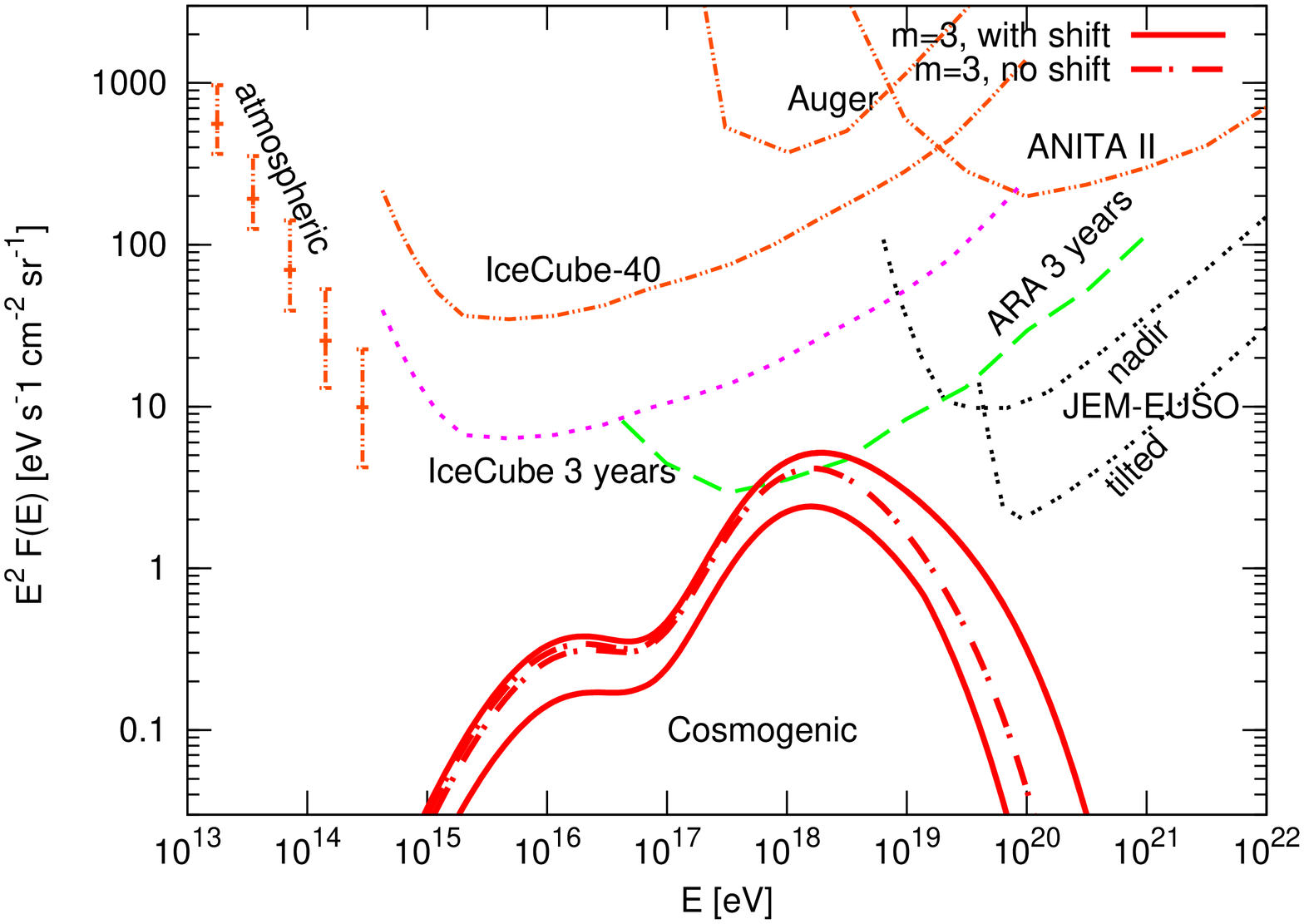}
\caption[...]{Maximum and minimum cosmogenic neutrino fluxes  averaged over flavors expected in the dip model for fixed values of the parameter $m$ in Eq.~\ref{evolution-function}. \ref{F7}.a (left panel)  For $m=2$ (blue lines) or $m= 4$ (red lines) shifting the energy scale of the HiRes spectrum. \ref{F7}.b (right panel) For $m=3$ and the VHE $\gamma$-ray spectrum of Ref.~\cite{Neronov:2011kg} only,  we show the range obtained by shifting the HiRes energy as in the left panel (larger range) and the  range obtained without any  energy shift. For $m=4$ no allowed range remains without an energy shift, thus we choose the larger fluxes  for which both ranges exist, which are for $m=3$.
}
\label{F7}
\end{figure}

\begin{figure}[ht]
\includegraphics[height=0.50\textwidth,clip=true,angle=270]{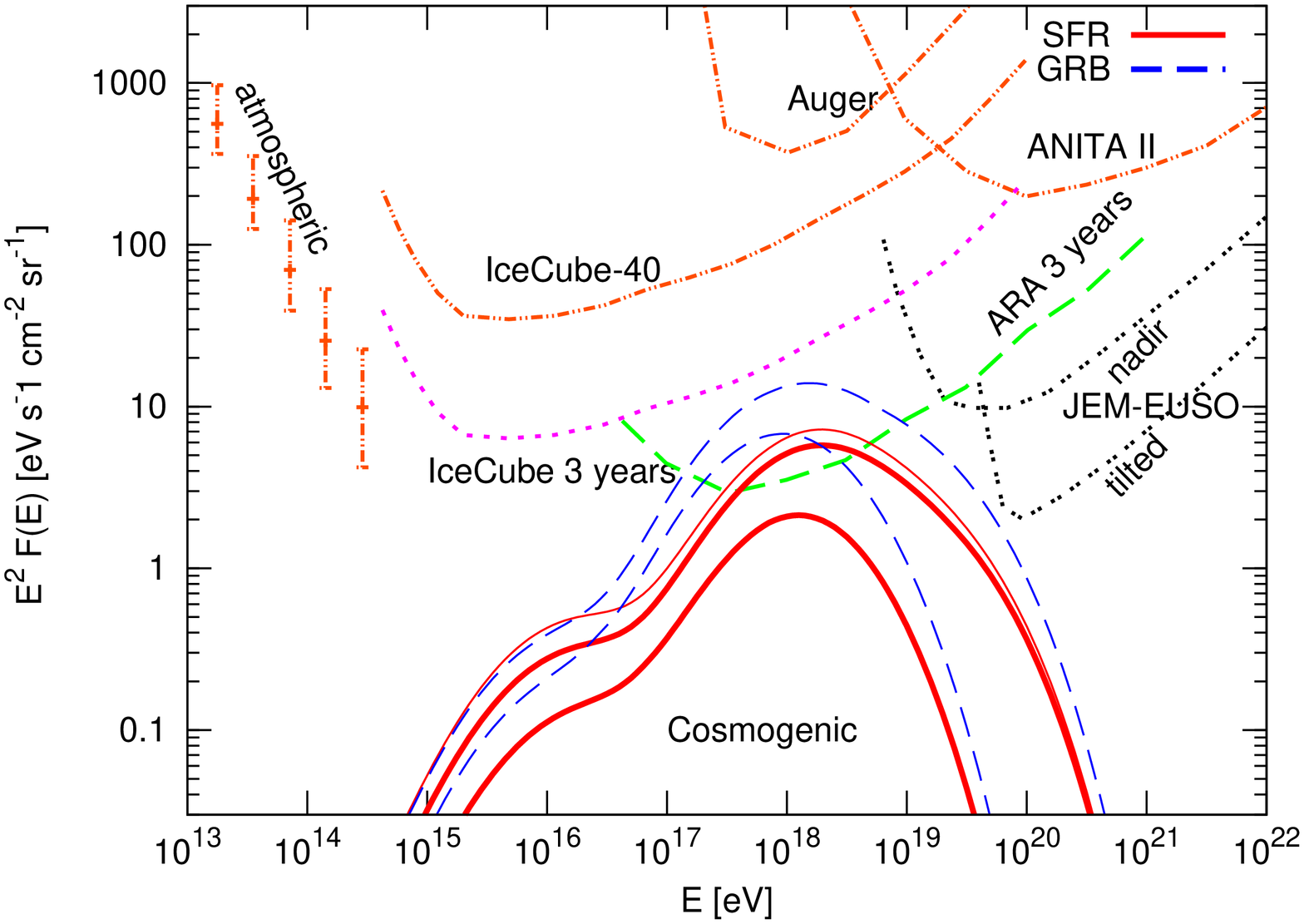}
\includegraphics[height=0.50\textwidth,clip=true,angle=270]{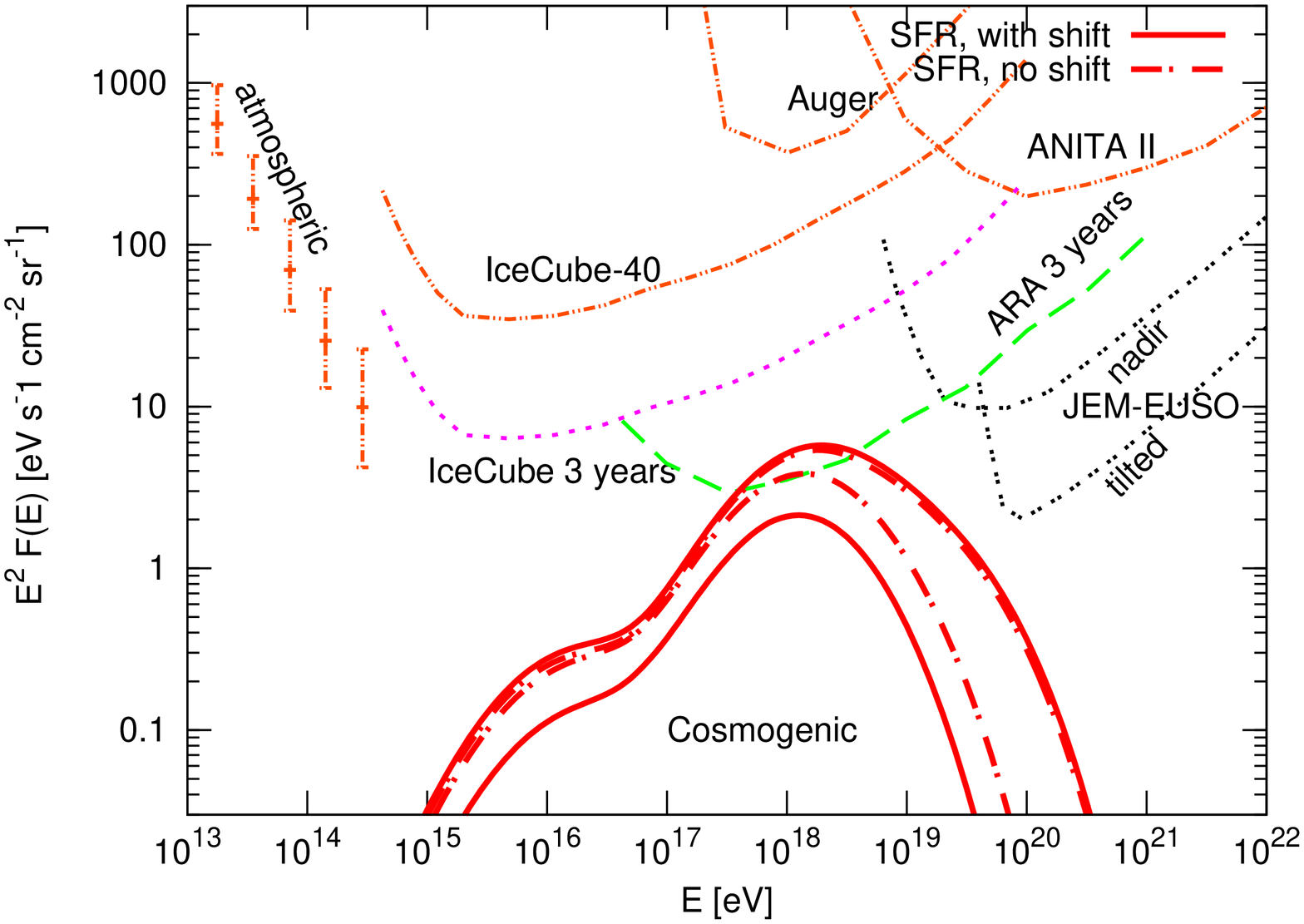}
\caption[...]{Maximum and minimum cosmogenic neutrino fluxes  averaged over flavors expected in the dip model assuming  the special evolution models in  Eq.~\ref{SFR}, \ref{GRB} and \ref{AGN} and varying $\alpha$ and $E_{\rm max}$. As in Figs.~\ref{F6} and \ref{F7}  in the  left panel (\ref{F8}.a)  the energy scale of the HiRes spectrum was shifted. Notice that dip models with the AGN evolution function in Eq.~\ref{AGN} are not allowed by any of the two VHE $\gamma$-ray backgrounds assumed and dip models with the GRB evolution in  Eq.~\ref{GRB} are rejected by the new lower $\gamma$-ray background estimate of Ref.~\cite{Neronov:2011kg}. \ref{F8}.b (right panel) For the SFR evolution of Eq.~\ref{SFR} and the VHE $\gamma$-ray spectrum of Ref.~\cite{Neronov:2011kg} only, we show the range obtained by shifting the energy scale of the HiRes spectrum as in the upper panel (larger range) and the  range obtained without any  energy shift.
}
\label{F8}
\end{figure}

As an example of our procedure in  Fig. \ref{F1} we show the chi-square $p$-value obtained in the first goodness of fit evaluation step described above of dip models with  $E_{\rm max}=10^{21}$~eV and $\alpha$  varied from 2.00 to 2.70 in 0.05 steps for the injected spectrum  and assuming the evolution function of Eq.~\ref{evolution-function} with $m$ varied from 0 to 5 in 0.5 steps. The $p$-values are shown in a color coded logarithmic scale, from best ($p=1$) to worse ($p$ close to zero). Only models with log($p$)$> -1.30$ are accepted in the following.  The only difference between the three panels in  Fig. \ref{F1}  is the upper bound on the  predicted VHE  $\gamma$-ray  fluxes  imposed: in   \ref{F1}.a-(upper panel)  no upper limit is imposed,  in   \ref{F1}.b-(middle panel) the First-Year Fermi spectrum is taken as  upper limit and in \ref{F1}.c-(lower panel) the spectrum of Ref.~\cite{Neronov:2011kg} is taken as  upper limit. Note that models with $m>4$ allowed in \ref{F1}.a are forbidden when the upper limit on the VHE photons fluxes is imposed in \ref{F1}.b. The more stringent constraint on the  diffuse $\gamma$-ray fluxes in \ref{F1}.c is more restrictive than the First-Year Fermi spectrum for $m \geq 3$. Notice in particular that the best fit models of the middle panel, with $m \geq 3.5$ and $\alpha$ between 2.4 and 2.5 are rejected in the right panel.  We can very easily see why this is so in  Fig. \ref{F4}, where we show the  predicted cosmic ray, gamma ray and neutrino  fluxes for the model with  $E_{\rm max}=10^{21}$~eV, $\alpha=2.35$ and $m=3.5$, which is the best (largest $p$-value)  model of Fig.~\ref{F1}.b  and is rejected in  Fig.~\ref{F1}.c. We see in  Fig. \ref{F4} that the predicted $\gamma$-ray spectrum saturates the First-Year Fermi diffuse spectrum (shown in black) but exceeds  the  lower background estimate of Ref.~\cite{Neronov:2011kg}. The flux of photons produced only through the GZK process  in this model is also shown in the figure (magenta dotted line). As in all models with $\alpha>2$, the dominant photon flux comes from $e^+ e^-$ pair production processes.
 Fig. \ref{F4} shows also the VHE $\gamma$-ray spectrum (magenta dot-dashed  line) that the model would predict if the older evaluation of the EBL of Ref.~\cite{Stecker:2005qs} (still valid in the scenario of Ref.~\cite{Essey:2009zg}) would be used. In this case the model would be somewhat less constrained by the upper bound on VHE photon fluxes. Besides, for maximal fluxes, the $\gamma$-rays above and below 10 GeV would require additional explanations. 

In  Figs.~\ref{F6}, \ref{F7} and \ref{F8}  we show the range of cosmogenic neutrino fluxes  averaged over  flavors as function of the energy expected in the dip model of UHECR when fixing one of the parameters of the model. 
 The fluxes shown with  thick lines correspond to the VHE $\gamma$-ray background evaluation of Ref.~\cite{Neronov:2011kg} and they are lower than the  fluxes shown with thin lines, which correspond to the First-Year Fermi VHE $\gamma$-ray spectrum. The results with the  First-Year Fermi VHE $\gamma$-ray spectrum are compared with those of the new lower evaluation of  Ref.~\cite{Neronov:2011kg} only in the left panels of  Figs.~\ref{F6}, \ref{F7} and \ref{F8}. Notice that in almost all cases only the maximum fluxes depend on the VHE $\gamma$-ray background assumed, because the lower neutrino fluxes are associated with VHE $\gamma$-ray fluxes much smaller that the upper bound. In this case there are only three lines for each model, two for maximum fluxes and one for the minimum fluxes. The range  of fluxes is obtained by choosing among all the models with $p> 0.05$ those with maximum and minimum neutrinos fluxes  at each energy, thus the models which provide the fluxes shown may change at each energy value.

 In the left panels of  Figs.~\ref{F6}, \ref{F7} and \ref{F8},  we take into account possible systematic errors in the HiRes energy scale, by shifting the energy scale by a factor in the range $(1.30)^{-1}$ to 1.30 (and the results of both VHE $\gamma$-ray backgrounds are shown). In the right panels of Figs.~\ref{F6}, \ref{F7} and \ref{F8} the ranges of neutrino fluxes with and without shifting the HiRes energy scale are compared for the particular case yielding the largest fluxes with the new VHE $\gamma$-ray 
spectrum of Ref.~\cite{Neronov:2011kg}. The  neutrino flux ranges obtained varying the energy are larger than with the energy fixed, but not by much. This is so because the HiRes spectrum without any energy shift is  very well fitted by the dip-model, so large changes in the HiRes spectrum do not provide a good fit. 
 
In Fig.~\ref{F6}.a (left panel) $E_{\rm max}$ is either fixed to $10^{20}$ eV  or varied within its assumed range, from $10^{20}$ eV  up to $10^{21}$ eV, and $\alpha$ as well as $m$ in the evolution function in Eq.~\ref{evolution-function} are varied as explained above  (and the HiRes energy is varied too).  Notice that for each energy there are only three lines, the highest corresponding to the maximum fluxes allowed by the First-Year Fermi $\gamma$-ray background. The maximum neutrino fluxes with the new $\gamma$-ray background evaluation of Ref.~\cite{Neronov:2011kg} are about a factor of 2 smaller. In Fig.~\ref{F6}.b  (right panel)  we compare  the neutrino flux range obtained by shifting the HiRes energy scale as presented in the left panel (larger range) with the  range obtained without any  energy shift, for  all  $E_{\rm max}$ up to $10^{21}$ eV and the VHE $\gamma$-ray spectrum of Ref.~\cite{Neronov:2011kg}. As Fig.~\ref{F6}.b shows,the difference in the ranges derived by allowing a variation of the HiRes energy scale and by kipping it fix is very small. 
Notice that the  models with $E_{\rm max}=10^{20}$ eV  in the left panel, are  not allowed when the HiRes energy scale  is  not shifted, as shown in the right panel.

 In Fig.~\ref{F7} the source evolution parameter $m$ in Eq.~\ref{evolution-function} is fixed and all other parameters varied. Models with $m= 0$ or 1 are disfavored (see e.g. Fig~\ref{F1}), and those with $m >4$ forbidden by the  new $\gamma$-ray upper limits,  thus neutrino flux ranges are presented in  Fig.~\ref{F7}.a (left panel) for $m=2$ and $m= 4$  (and the HiRes energy varied as explained above). In  \ref{F7}.b (right panel) we compare the range obtained by shifting the HiRes energy as in the left panel (larger range) with the  range obtained without any  energy shift for $m=3$ and the VHE $\gamma$-ray spectrum of Ref.~\cite{Neronov:2011kg} only.  For $m=4$, the case with larger fluxes in the left panel, no allowed range remains without an energy shift. Thus we present  in  \ref{F7}.b the case  with the largest neutrino fluxes  for which both ranges (i.e. with and without energy shift) exist, which is  the case of  $m=3$.
  
  In Fig.~\ref{F8} one of the specific  source evolution functions in Eqs.~\ref{SFR}, \ref{GRB} and \ref{AGN}  is chosen for each range and $E_{\rm max}$ and $\alpha$ are varied as explained before.  Notice that these evolution functions were not included in the previous two figures.   
   Fig.~\ref{F8}.a (left panel) shows that dip models with the AGN evolution function of Eq.~\ref{AGN} are rejected  by both VHE $\gamma$-ray background evaluations we take as upper bounds to the gamma-ray fluxes produced and dip models with the GRB evolution in Eq.~\ref{GRB} are not allowed by the lower VHE $\gamma$-ray spectrum of Ref.~\cite{Neronov:2011kg}. Thus, Fig.~\ref{F8}.a  shows the range of neutrino fluxes for the SFR evolution function in Eq.~\ref{SFR} for both VHE  $\gamma$-ray spectra. Only the maximum fluxes change with the VHE $\gamma$-ray spectrum. The minimum fluxes are independent of the upper limit provided by VHE $\gamma$-rays.  The range, which is higher, corresponding to the GRB evolution is presented only for the  First-Year Fermi VHE $\gamma$-ray spectrum because no range remains. 
   
   As in the left panels of the previous two figures,  Fig.~\ref{F8}.a shows the maximum and minimum cosmogenic neutrino fluxes obtained by also varying the HiRes energy scale. In \ref{F8}.b (right panel)  we compare the range obtained by shifting the energy scale of the HiRes spectrum as in the left panel (larger range) with the  range obtained without any  energy shift only for the SFR evolution of Eq.~\ref{SFR}, the only one of the three considered allowed by the VHE $\gamma$-ray spectrum of Ref.~\cite{Neronov:2011kg} which is assumed in this panel.

\begin{figure}[ht]                
\begin{center}
\includegraphics[height=0.50\textwidth,clip=true,angle=270]{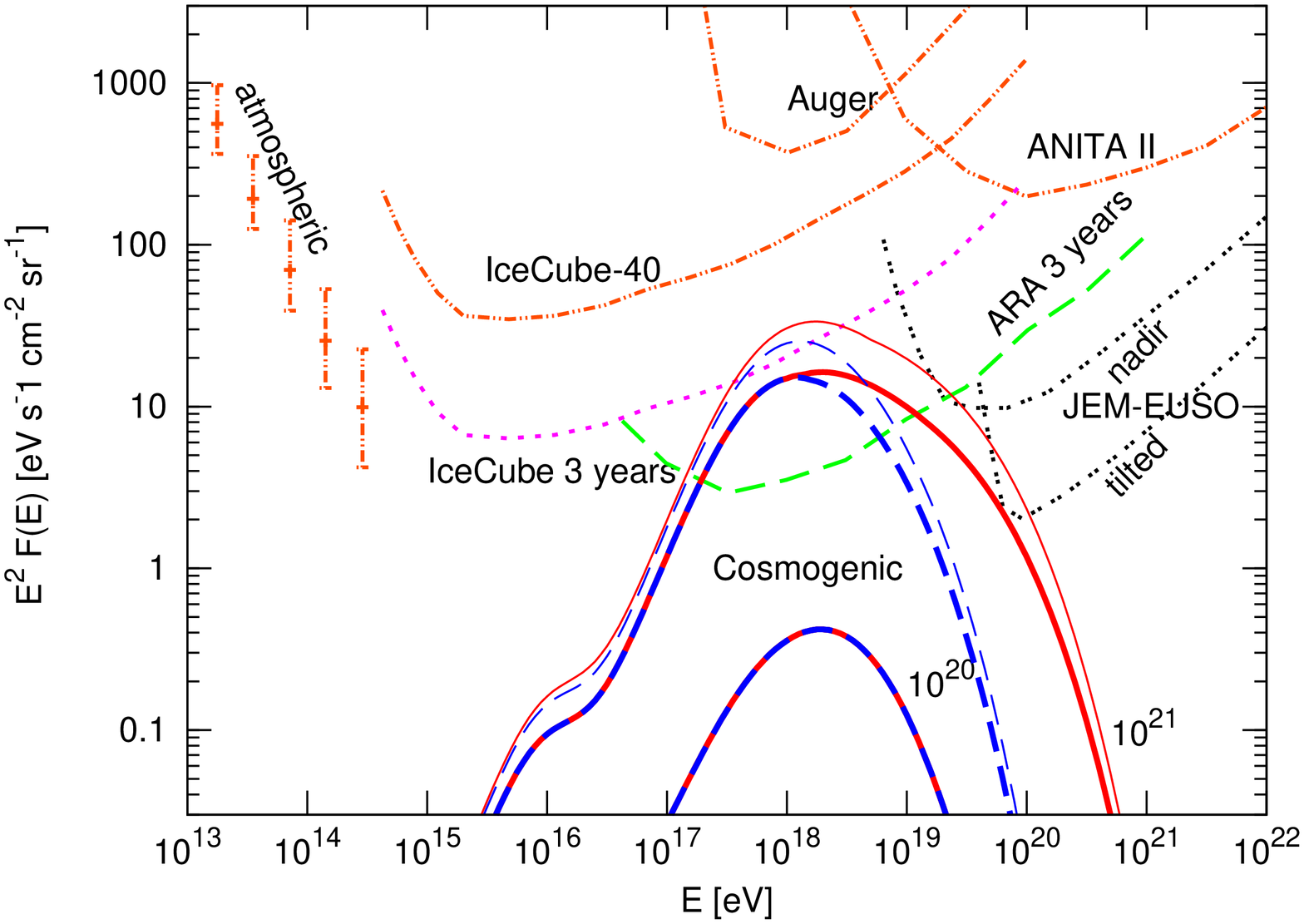}
\includegraphics[height=0.50\textwidth,clip=true,angle=270]{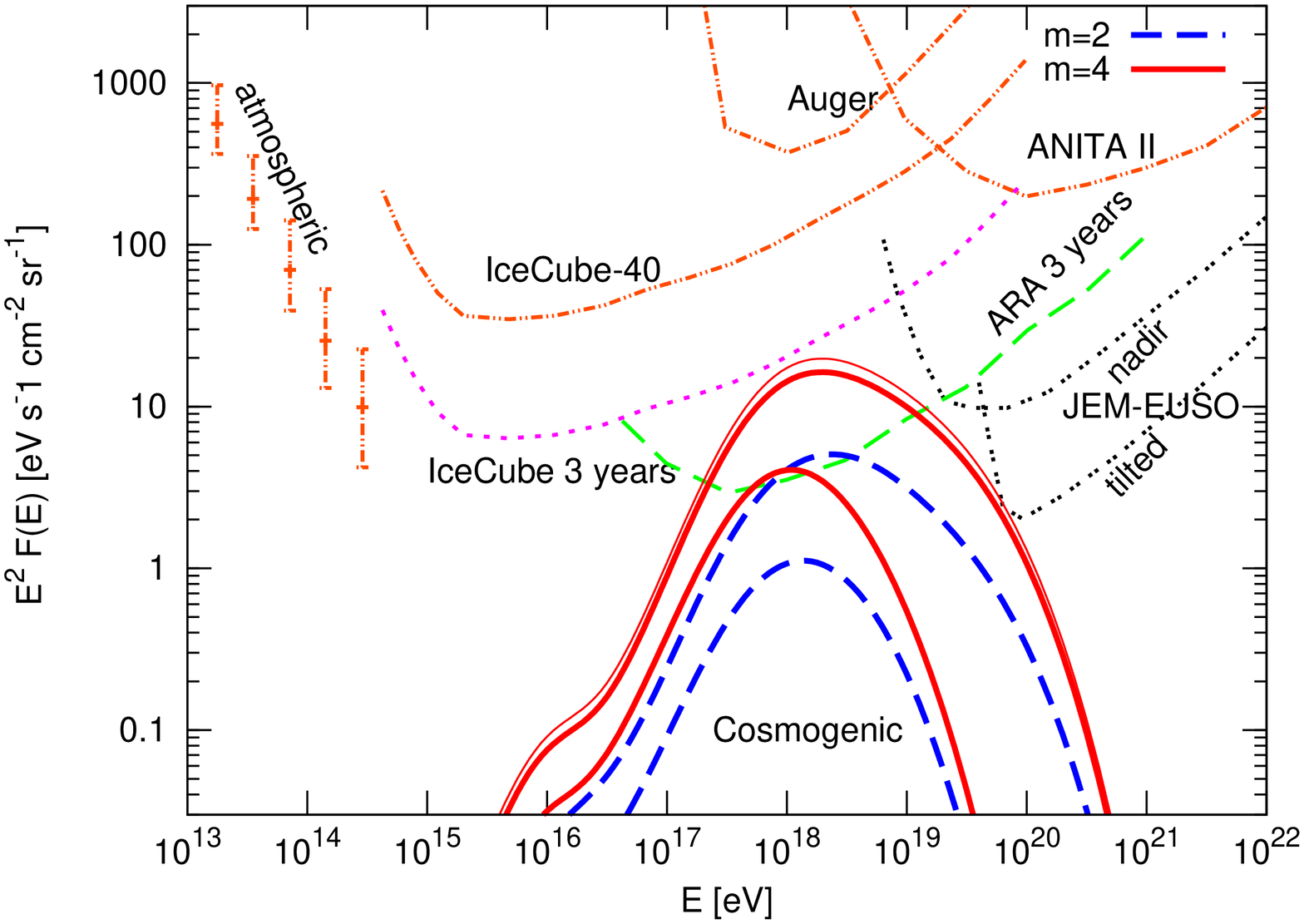}
\includegraphics[height=0.50\textwidth,clip=true,angle=270]{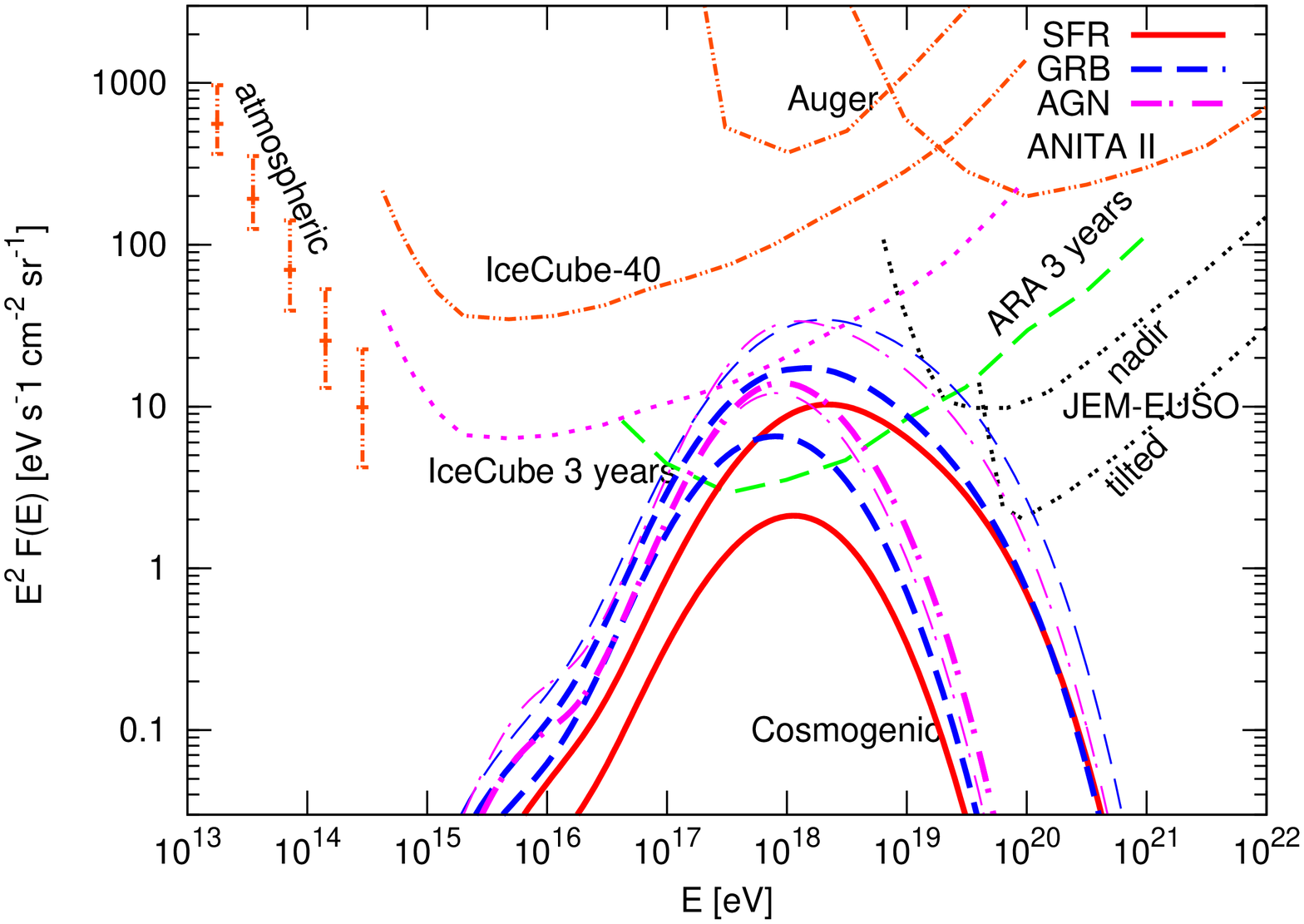}
\end{center}
\caption[...]{Upper, middle and lower panels similar respectively to Figs.~\ref{F6}.a, \ref{F7}.a and \ref{F8}.a but fitting the HiRes spectrum only above 10$^{19}$ eV  (and $E_{\rm min} =10^{18}$ eV).
}
\label{F9}
\end{figure}

  Figs.~\ref{F6}, \ref{F7} and \ref{F8}   also show the atmospheric neutrino flux measured by  IceCube~\cite{IceCube_atmosphere}, existing  experimental upper limits  from Auger~\cite{Auger_nu_2009},
 ANITA II~\cite{ANITAII}  and IceCube-40 and expected sensitivity regions for IceCube in 3 years~\cite{IceCube40}, JEM-EUSO~\cite{JEM-EUSO-neutrinos} and ARA~\cite{ARA-neutrinos}.  The JEM-EUSO bound corresponds to 1 event per energy decade per year. All the others are 90\% CL bounds according to the respective references.  The differential flux bound corresponding to  3 years of observation of IceCube was not provided by the IceCube collaboration. It was obtained by scaling down the IceCube-40 differential flux bound by the ratio of the integrated limits (which assume an $E^{-2}$ neutrino spectrum) shown in Fig.~3 of Ref.~\cite{IceCube40}. This procedure does not take into account the presence of the atmospheric neutrino background and thus it is not correct at low energies, but it should be fine in the energy region where the cosmogenic neutrino fluxes are maximal. Notice that of all the expected sensitivities shown only that corresponding to 3 years of observation with ARA enter into the cosmogenic neutrino ranges shown.  The maximum cosmogenic neutrino fluxes are a factor of 2 or 3 below the sensitivity of IceCube in 3 years. Thus at least about 10 years of observation with IceCube or a similar KM3NeT detector would be needed  for detection.    

 \section{Range of Cosmogenic Neutrino Fluxes in the Ankle Model of UHECR}  
 
  Because we are interested in producing upper bounds to the cosmogenic neutrinos fluxes, we also consider what in Ref.~\cite{Berezinsky:2010xa} is called the ``ankle model'', in which we fit with protons the HiRes spectrum  only above $E_{\rm fit} = 10^{19}$ eV. This is not a self consistent model, since it does not explain the UHECR data below this energy. An additional component of UHECR needs to be introduced. The nature of this extra component will dictate if there is an additional contribution to the $\gamma$-ray flux. If  it is dominated by extragalactic protons, we will have a more complicated version of a ``dip"-like model, in which
  there would be a large contribution to the  $\gamma$-ray background and, as a result, lower allowed  maximum neutrino fluxes. Galactic protons would contradict the isotropy of the observed UHECR flux. The only other possibility are heavier nuclei, either galactic or extragalactic. But this would contradict the composition measurements of both HiRes and Auger~\cite{chem_HiRes, xmax}. 
  
   We assume in this model that the cutoff energy in the emission spectrum (see Eq.\ref{proton_flux}) is $E_{\rm min} = 10^{18}$ eV. The neutrino flux ranges are then obtained as explained in the previous section. The ankle model was also studied in Refs.~\cite{Berezinsky:2010xa}  and \cite{Ahlers:2010fw} and found to produce larger neutrino fluxes.
 
 The range of fluxes are shown in Fig.~\ref{F9}. The three panels of Fig.~\ref{F9} are similar to Figs.~\ref{F6}.a, \ref{F7}.a and \ref{F8}.a except for the higher values of  $E_{\rm min}$ and $E_{\rm fit}$. In Fig.~\ref{F9}.a (upper panel) $E_{\rm max}$ is fixed and $\alpha$ as well as $m$ in the evolution function in Eq.~\ref{evolution-function} are varied as explained above. In Fig.~\ref{F9}.b (middle  panel) the source evolution parameter $m$ in Eq.~\ref{evolution-function} is fixed and all other parameters varied. In Fig.~\ref{F9}.c (lower  panel)   one of the specific  source evolution functions in Eqs.~\ref{SFR}, \ref{GRB} and \ref{AGN}  is chosen for each range and $E_{\rm max}$ and $\alpha$ are varied as explained before.  In all three panels the HiRes energy has been  varied as explained above.

 Fig.~\ref{F9}.c shows that in the ankle model,  both the GRB and AGN  source evolution functions are allowed, even with the lower $\gamma$-ray background we used. For the AGN evolution, only a very small range of parameters are allowed,  so that the maximum and minimum neutrino fluxes (shown in thick magenta dot-dashed lines) practically coincide.

 The maximum neutrino fluxes  are here a factor of 2 to 3 larger than in the dip model, which puts them within the reach of IceCube in three years of observation.

\section{Conclusions}

Cosmogenic neutrinos constitute one of the main high energy signals expected in neutrino telescopes, such as IceCube~\cite{ICECUBE},  ANITA~\cite{ANITA},  the future KM3NeT~\cite{KM3NeT} and ARA~\cite{ARA} or space based observatories such as JEM-EUSO~\cite{JEM-EUSO}.  They could also be observed by Pierre Auger Observatory~\cite{Auger} and the Telescope Array~\cite{TA}. 
Here we present the largest neutrino fluxes compatible with cosmic rays and gamma-rays observations.

 Proton primaries produce larger maximum cosmogenic neutrino fluxes than heavier nuclei. Thus we consider the  dip model, which assumes that all UHECR primaries above 10$^{18}$ eV are protons  and is consistent with both the spectrum and composition data of HiRes.  
 As it  has been  already pointed out~\cite{Berezinsky:2010xa, Ahlers:2010fw} 
fitting the UHECR spectrum  with protons only above  higher energies, such as 10$^{19}$ eV, leads to higher neutrino fluxes. This, however, is an incomplete model of UHECR unless a consistent explanation of the dominant component below this energy is presented. With this caveat, we  also consider this model, which following Ref.~\cite{Berezinsky:2010xa} we call the ankle model, to obtain the highest allowed cosmogenic neutrino fluxes.

We use a program which allows for fast analytic calculations of the secondary proton, photon and neutrino spectrum to find at each energy the expected range of cosmogenic neutrino fluxes. We require the produced UHECR spectrum to fit the HiRes spectrum at energies above a certain energy $E_{\rm fit}$, which is 10$^{18}$ eV in the dip model and 10$^{19}$ eV in the ankle model. Below this energy, the HiRes spectrum is taken  as an upper bound. We also impose as an upper bound to the  predicted gamma-ray fluxes  the new evaluation of the diffuse extragalactic VHE $\gamma$-ray background of Ref.~\cite{Neronov:2011kg} and compare the results with those derived assuming instead the First-Year Fermi~\cite{Abbasi:2007sv} background, which is higher than the former by about a factor of 2 at energies above 10 GeV.

The resulting neutrino flux ranges are presented in Figs.~\ref{F6} to \ref{F9} together with experimental bounds of IceCube-40, Auger and ANITA and sensitivity regions  of  IceCube in 3 years, ARA in 3 years and JEM-EUSO. 

 In the left panels of Figs.~\ref{F6}, \ref{F7} and \ref{F8}  we show the range of cosmogenic neutrino fluxes as function of the energy expected in the dip-model of UHECR allowing for a shift of the HiRes energy scale by a factor between $(1.30)^{-1}$ and 1.30, to take into account possible systematic errors. Also in these panels the results obtained by taking as upper bounds both VHE $\gamma$-ray backgrounds are compared.  The  comparison of the ranges of neutrino fluxes obtained in the dip model with and without shifting the nominal HiRes energy are presented in the right  panels of Figs.~\ref{F6}, \ref{F7} and \ref{F8}  for the particular case yielding the largest fluxes with the new VHE $\gamma$-ray spectrum of Ref.~\cite{Neronov:2011kg} in the corresponding left panel of the same figure. The ranges obtained varying the energy are larger than with the energy fixed, but not by much, because the HiRes spectrum without any energy shift is very well fitted by the dip model, so large changes in the HiRes spectrum do not provide good fits.
 
 In  Figs.~\ref{F6} and \ref{F7}, the evolution model in Eq.~\ref{evolution-function} is used.  In these figures, the highest cosmogenic neutrino fluxes allowed by the lower estimate of the VHE $\gamma$-ray background~\cite{Neronov:2011kg} are about a factor of 2 lower that those obtained by imposing the First-Year Fermi background.   
 
 Figs.~\ref{F6}, \ref{F7} and \ref{F8} show that  detecting the highest cosmogenic neutrino fluxes predicted by the dip model would require about 10 years of observation with IceCube (or a similar KM3NeT) or the construction of the new ARA detector.

The upper and middle panels of Fig.~\ref{F9} are similar to the left panels of Figs.~\ref{F6} and \ref{F7} but apply to the  ankle model. Here the maximum cosmogenic neutrino fluxes are a factor of 2 to 3 higher than in the dip model, what brings them into the range of detectability of IceCube in 3 years.  
Our conclusions are consistent with those of Ref.~\cite{Ahlers:2010fw}  in that the ankle model produces maximal neutrino fluxes which could be detected with a few years of observation of IceCube, and are not consistent with those of Ref.~\cite{Berezinsky:2010xa} in this regard.

Refs.~\cite{Berezinsky:2010xa} and \cite{Ahlers:2010fw} use the evolution model in Eq.~\ref{evolution-function}. We also consider here the special evolution functions  in Eqs.~~\ref{SFR}, \ref{GRB} and \ref{AGN}, in which the UHECR sources are assumed to have the same evolution of either the star formation rate (SFR),  or the gamma-ray burst (GRB) rate, or the active galactic nuclei (AGN) rate in the Universe.  To our knowledge the limits imposed by the diffuse extragalactic VHE $\gamma$-ray background extracted from Fermi-LAT data on these evolution functions had not been studied before.

Fig.~\ref{F8} shows that the new  VHE $\gamma$-ray background~\cite{Neronov:2011kg} rejects the dip models assumed to have the GRB source evolution function in Eq.~\ref{GRB}. The dip models with the AGN source evolution function in Eq.~\ref{AGN} are rejected by any of the two VHE $\gamma$-ray backgrounds, so models with this evolution do not appear in the figure. Only dip models with the SFR evolutions function
in Eq.~\ref{SFR} is allowed by both observed  VHE $\gamma$-ray backgrounds assumed, and both ranges (only the maximum fluxes change) are very similar. The lower panel of Fig.~\ref{F9}  shows that in the ankle model,  both the GRB and AGN  source evolution functions are allowed, even with the lower $\gamma$-ray background we used (in the case of the AGN evolution barely so).

\vspace{0.3cm}
{\bf Acknowledgments}
The numerical calculations in this work were performed with the Computational Cluster of the Theoretical Division of INR RAS. G.G was supported in part by the US DOE grant DE-FG03-91ER40662 Task C.
 O.K. was supported in part by the RFBR grant 10-02-01406-a. G.G. wants to thank the APC, Paris, for its hospitality  during part of the period in which this paper was written.


\begin{thebibliography}{99}

\bibitem{Auger} The Pierre Auger Observatory, http://www. auger.org.

\bibitem{xmax}
  J.~Abraham {\it et al.}  [Pierre Auger Collaboration],
  Phys.\ Rev.\ Lett.\  {\bf 104}, 091101 (2010);
  [arXiv:1002.0699 [astro-ph.HE]].
%
  
  \bibitem{correlation} 
  J.~Abraham {\it et al.}  [Pierre Auger Collaboration],
  Science {\bf 318}, 938 (2007)
  [arXiv:0711.2256 [astro-ph]];
  J.~Abraham {\it et al.}  [Pierre Auger Collaboration],
  Astropart.\ Phys.\  {\bf 29}, 188 (2008)
  [Erratum-ibid.\  {\bf 30}, 45 (2008)]
  [arXiv:0712.2843 [astro-ph]];
  P.~Abreu {\it et al.}  [Pierre Auger Observatory Collaboration],
  Astropart.\ Phys.\  {\bf 34}, 314 (2010)
  [arXiv:1009.1855 [Unknown]].
  
  \bibitem{Semikoz-Virgo} 
  G.~Giacinti, M.~Kachelriess, D.~V.~Semikoz and G.~Sigl,
  JCAP {\bf 1008}, 036 (2010)
  [arXiv:1006.5416 [astro-ph.HE]];
  D.~V.~Semikoz,
  arXiv:1009.3879 [astro-ph.HE];
  G.~Giacinti and D.~V.~Semikoz,
  Phys.\ Rev.\  D {\bf 83}, 083002 (2011)
  [arXiv:1011.6333 [astro-ph.HE]].
  
  
  \bibitem{HiRes} 
  The High Resolution  Fly's Eye Collaboration (HiRes),
    http://hires.physics.utah.edu/
  
\bibitem{Abbasi:2009nf}
  J.~Belz  [HiRes Collaboration],
  Nucl.\ Phys.\ Proc.\ Suppl.\  {\bf 190} (2009) 5.
  R.~U.~Abbasi {\it et al.}  [HiRes Collaboration],
  Phys.\ Rev.\ Lett.\  {\bf 104}, 161101 (2010)
  [arXiv:0910.4184 [astro-ph.HE]].


 \bibitem{chem_HiRes} 
 R.~U.~Abbasi {\it et al.}  [The High Resolution  Fly's Eye Collaboration], 
  Astrophys.\ J.\  {\bf 622}, 910 (2005)
   [arXiv:astro-ph/0407622];
  D.~R.~Bergman {\it et al.}, 
  ``UHECR composition measurements using
  the HiRes-II detector,'' Proc. 29th ICRC (Pune), 2005
  [astro-ph/0507483].


\bibitem{Abbasi:2010xt}
  R.~U.~Abbasi {\it et al.},
  Astropart.\ Phys.\  {\bf 30}, 175 (2008)
  [arXiv:0804.0382 [astro-ph]] and
  arXiv:1002.1444 [astro-ph.HE]. 
    %



 \bibitem{dip-model}
 C.~T.~Hill and D.~N.~Schramm, 
Phys.\ Rev.\  D {\bf 31}, 564 (1985); 
V.~S.~Berezinsky and S.~I.~Grigorieva,
Astron. Astrophys. {\bf199} (1988) 1;
  V.~Berezinsky, A.~Z.~Gazizov and S.~I.~Grigorieva, 
 Phys.\ Rev.\ D {\bf 74}, 043005 (2006) [arXiv:hep-ph/0204357]; 
astro-ph/0210095;
Nucl.\ Phys.\ Proc.\ Suppl.\  {\bf 136}, 147 (2004) [astro-ph/0410650];
Phys.\ Lett.\ B {\bf 612} (2005) 147 [astro-ph/0502550].


\bibitem{minimal-GKS} 
  G.~Gelmini, O.~Kalashev and D.~V.~Semikoz,
  Astropart.\ Phys.\  {\bf 28}, 390 (2007)
  [arXiv:astro-ph/0702464] and
  arXiv:0706.3847 [astro-ph].


\bibitem{Abbasi:2007sv}
  R.~U.~Abbasi {\it et al.}  [HiRes Collaboration],
  Phys.\ Rev.\ Lett.\  {\bf 100}, 101101 (2008)
  [arXiv:astro-ph/0703099].


\bibitem{PAO-spectrum}
  J.~Abraham {\it et al.}  [Pierre Auger Collaboration],
  Phys.\ Rev.\ Lett.\  {\bf 101}, 061101 (2008)
  [arXiv:0806.4302 [astro-ph]].
  J.~Abraham {\it et al.}  [The Pierre Auger Collaboration],
  Phys.\ Lett.\  B {\bf 685}, 239 (2010)
  [arXiv:1002.1975 [astro-ph.HE]].
  
  \bibitem{gzk}
K.~Greisen,
Phys.\ Rev.\ Lett.\  {\bf 16}, 748 (1966).
G.~T.~Zatsepin and V.~A.~Kuzmin,
JETP Lett.\  {\bf 4}, 78 (1966)
[Pisma Zh.\ Eksp.\ Teor.\ Fiz.\  {\bf 4}, 114 (1966)].


\bibitem{bere} V. S. Berezinsky and G. T. Zatsepin, Phys. 
Lett {\bf 28B}, 423 (1969).

\bibitem{GKS} 
  G.~Gelmini, O.~E.~Kalashev and D.~V.~Semikoz,
  J.\ Exp.\ Theor.\ Phys.\  {\bf 106}, 1061 (2008)
  [arXiv:astro-ph/0506128];
  G.~B.~Gelmini, O.~E.~Kalashev and D.~V.~Semikoz,
  JCAP {\bf 0711}, 002 (2007)
  [arXiv:0706.2181 [astro-ph]].

\bibitem{augersdphotonfractionlimit}
  J.~Abraham {\it et al.}  [Pierre Auger Collaboration],
  Astropart.\ Phys.\  {\bf 29}, 243 (2008).
  [arXiv:0712.1147 [astro-ph]].
  J.~Abraham {\it et al.}  [The Pierre Auger Collaboration],
  Astropart.\ Phys.\  {\bf 31}, 399 (2009);
  [arXiv:0903.1127 [astro-ph.HE]].

\bibitem{reviewGZKneutrinos}
O.~E.~Kalashev, V.~A.~Kuzmin, D.~V.~Semikoz and G.~Sigl,
Phys.\ Rev.\ D {\bf 66}, 063004 (2002)
[arXiv:hep-ph/0205050].

\bibitem{Semikoz:2003wv}
D.~V.~Semikoz and G.~Sigl,
JCAP {\bf 0404}, 003 (2004)
[arXiv:hep-ph/0309328].

\bibitem{ICECUBE} IceCube, http://icecube.wis.edu/.

\bibitem{ANITA} Antarctic Impulse Transient Array (ANITA),
  http:// www.ps.uci.edu/anita/.

\bibitem{KM3NeT} Cubic Kilometre Neutrino Telescope (KM3NeT), http:// www.km3net.org/home.php

\bibitem{ARA}  Askaryan Radio Array (ARA), http://ara.physics.wisc. edu/

\bibitem{JEM-EUSO} Extreme Universe Space Observatory onboard  the  Japa nese Experiment Module (JEM-EUSO),
http://jemeuso. riken.jp/en/index.html

\bibitem{TA} Telescope Array (TA),  http://www.telescopearray.org/

\bibitem{HiRes-spectrum}
http://www.physics.rutgers.edu/\%7Edbergman/HiRes-Monocular-Spectra-200702.html

\bibitem{Abdo:2010nz}
  A.~A.~Abdo {\it et al.}  [The Fermi-LAT collaboration],
  Phys.\ Rev.\ Lett.\  {\bf 104}, 101101 (2010)
  [arXiv:1002.3603 [astro-ph.HE]].

\bibitem{Neronov:2011kg}
  A.~Neronov and D.~V.~Semikoz,
  arXiv:1103.3484 [astro-ph.CO].

\bibitem{wdowczyk}
J. Wdowczyk , W. Tkaczyk, C. Adcock and A. W. Wolfendale, 
J. Phys. A: Gen. Phys {\bf 4}
L37-9 (1971); J. Wdowczyk , W. Tkaczyk and A. W. Wolfendale,
 J. Phys. A: Gen. Phys {\bf 5} 1419 (1972);
V. Berezinsky and  A. Yu. Smirnov, Astrophys. Sp. Sci. {\bf 32}, 461  (1975);
 J. Wdowczyk  and A. W. Wolfendale, Astrophys. 
Jour. {\bf 349}, 35 (1990);

\bibitem{Berezinsky:2010xa}
  V.~Berezinsky, A.~Gazizov, M.~Kachelriess and S.~Ostapchenko,
  Phys.\ Lett.\  B {\bf 695}, 13 (2011)
  [arXiv:1003.1496 [astro-ph.HE]].
  
\bibitem{Ahlers:2010fw}
  M.~Ahlers, L.~A.~Anchordoqui, M.~C.~Gonzalez-Garcia, F.~Halzen and S.~Sarkar,
  Astropart.\ Phys.\  {\bf 34}, 106 (2010)
  [arXiv:1005.2620 [astro-ph.HE]].

\bibitem{kks1999} 
O.E.~Kalashev, V.A.~Kuzmin and D.V.~Semikoz,
astro-ph/9911035;
   Mod.\ Phys.\ Lett.\ A {\bf 16}, 2505 (2001) [astro-ph/0006349].
O.E.~Kalashev Ph.D. Thesis, INR RAS, 2003.

\bibitem{reviews1} P.~Bhattacharjee, G.~Sigl, Phys. Rept. {\bf 327},
109 (2000).

\bibitem{Franceschini2008} 
A.~Franceschini, G.~Rodighiero and M.~Vaccari, 
Astron. \& Astrophys. {\bf 487} 837 (2008) [arXiv:0805.1841]. 

\bibitem{Kneiske2010} 
T.~M.~Kneiske and H.~Dole, 
Astron. \& Astrophys.  {\bf 515} A19 (2010) [arXiv:1001.2132]. 

\bibitem{Dominguez2010} 
A.~Dominguez {\it et al.}, 
MNRAS {\bf 410} 2556 (2010) [arXiv:1007.1459]. 

\bibitem{Dolag:2004kp}
  K.~Dolag, D.~Grasso, V.~Springel and I.~Tkachev,
  JCAP {\bf 0501}, 009 (2005).
  [astro-ph/0410419].
  
\bibitem{Neronov:2009gh}
  A.~Neronov and D.~V.~Semikoz,
  Phys.\ Rev.\  D {\bf 80}, 123012 (2009)
  [arXiv:0910.1920 [astro-ph.CO]].

\bibitem{Aharonian} 
X.~Wang, R.~Liu, F.~Aharonian,
 arXiv:1103.3574 [astro-ph.HE]
 
 \bibitem{Yuskel-2008}
H. Y\"{u}ksel, M. D. Kistler, J. F.  Beacom and A. M.  Hopkins,
Ap. J. {\bf 638} L5 (2008).

\bibitem{Yuskel-2007}
H. Y\"{u}ksel and M.D. Kistler
Phys.\ Rev.\ D {\bf 75}, 083004 (2007).

\bibitem{Hasinger-2005}
G. Hasinger, T. Miyaji, M. Schmidt,  Astron. and Astroph. {\bf 441} 417 (2005)

\bibitem{Ahlers:2009rf}
  M.~Ahlers, L.~A.~Anchordoqui and S.~Sarkar,
  Phys.\ Rev.\  D {\bf 79}, 083009 (2009)
  [arXiv:0902.3993 [astro-ph.HE]].
  
\bibitem{Kachelriess:2004pc}
  M.~Kachelriess and D.~Semikoz,
  Astropart.\ Phys.\  {\bf 23}, 486 (2005)
  [arXiv:astro-ph/0405258];
  M.~Kachelriess, S.~Ostapchenko and R.~Tomas,
  New J.\ Phys.\  {\bf 11}, 065017 (2009)
  [arXiv:0805.2608 [astro-ph]].
  
\bibitem{Kneiske:2003tx}
  T.~M.~Kneiske, T.~Bretz, K.~Mannheim and D.~H.~Hartmann,
  Astron.\ Astrophys.\  {\bf 413}, 807 (2004)
  [arXiv:astro-ph/0309141].
  
\bibitem{statistics}  
S. Baker and R.D. Cousins,
Nucl. Instrum. Methods {\bf221}, 437 (1984); Particle Data Group's
Statistics Review (2004).

\bibitem{Stecker:2005qs} 
F.~W.~Stecker, M.~A.~Malkan and S.~T.~Scully,
  Astrophys.\ J.\  {\bf 648}, 774 (2006)
  [arXiv:astro-ph/0510449].
  
\bibitem{Essey:2009zg}
  W.~Essey and A.~Kusenko,
  Astropart.\ Phys.\  {\bf 33}, 81 (2010)
  [arXiv:0905.1162 [astro-ph.HE]];
  W.~Essey, O.~E.~Kalashev, A.~Kusenko and J.~F.~Beacom,
  Phys.\ Rev.\ Lett.\  {\bf 104}, 141102 (2010)
  [arXiv:0912.3976 [astro-ph.HE]].

\bibitem{IceCube_atmosphere}
   R.~Abbasi {\it et al.}  [IceCube Collaboration],
   arXiv:1104.5187 [astro-ph.HE].
        
 \bibitem{Auger_nu_2009}
  J.~Abraham {\it et al.}  [Pierre Auger Collaboration],
  Phys.\ Rev.\  D {\bf 79}, 102001 (2009)
  [arXiv:0903.3385 [astro-ph.HE]].

\bibitem{ANITAII}
   P.~W.~Gorham {\it et al.}  [The ANITA Collaboration],
   Phys.\ Rev.\  D {\bf 82}, 022004 (2010)
   [arXiv:1003.2961 [astro-ph.HE]];
  Erratum:  arXiv:1011.5004 [astro-ph.HE].
   
   \bibitem{IceCube40}
   R.~Abbasi {\it et al.}  [IceCube Collaboration],
   Phys.\ Rev.\  D {\bf 83}, 092003 (2011)
   [arXiv:1103.4250 [astro-ph.CO]].
   
   \bibitem{JEM-EUSO-neutrinos}
N. Inoue, K. Miyazawa and Y. Kawasaki [JEM-EUSO Collaboration],
Nucl. Phys. B (Proc. Suppl.) {\bf 196} 135 (2009).

\bibitem{ARA-neutrinos}
   P.~Allison {\it et al.},
   arXiv:1105.2854 [astro-ph.IM].

\end{thebibliography}
\end{document}